\documentclass[12pt]{article}

\def\lromn#1{\uppercase\expandafter{\romannumeral#1}}

\def\blist{\begin{list}{\setlength{\rightmargin}{\leftmargin}}}
\def\elist{\end{list}}
\usepackage{graphicx}

\begin{document}

\begin{flushright}
 ICRR-Report-499-2003-3
\end{flushright}

\begin{center}
\begin{large}

\textbf{
 Time Evolution of Tunneling in Thermal Medium\\
 -- Environment-driven Excited Tunneling --
 }

\end{large}

\vspace{2cm}

\begin{large}
Sh. Matsumoto and M. Yoshimura

Institute for Cosmic Ray Research, 
University of Tokyo \\
Kashiwa-no-ha 5-1-5 Kashiwa, Chiba \\
277-8582 Japan
\end{large}

\vspace{4cm}

{\bf ABSTRACT}
\end{center}

Time evolution of tunneling phenomena proceeding in thermal medium is
studied using a standard model of environment interaction. A
semiclassical probability formula for the particle motion in a
metastable state of one dimensional system put in thermal medium is
combined with the formula of quantum penetration factor through a
potential barrier, to derive the tunneling rate in medium. Effect of
environment, its influence on time evolution in particular, is clarified
in a real-time formalism. A nonlinear resonance effect is shown to
enhance the tunneling rate at finite times of order $2/\eta $, with
$\eta $ the friction coefficient. In the linear approximation this
effect has relevance to the parametric resonance. This effect enhances
the possibility of early termination of the cosmological phase
transition much prior to the typical Hubble time.

\newpage

\lromn 1 \hspace{0.2cm} {\bf Introduction}
\vspace{0.5cm} 

Tunneling phenomena, when they occur in isolation, 
are genuinely a quantum effect.
But when they occur in some surrounding medium,
the important question arises as to whether the tunneling
rate is enhanced or suppressed by the environment effect.
There are already many works on this subject
\cite{q-tunneling review1}, and
most past works deal with a system in equilibrium as a whole.
The Euclidean technique such as the bounce solution
\cite{bounce} is often used in this context
\cite{caldeira-leggett 83}-\cite{zweger}.
We would like to examine this problem by working in a new
formalism of the real-time description of tunneling 
phenomena \cite{my 00-1}-\cite{my 00-3}.
We find it more illuminating to use a real-time
formalism instead of the Euclidean method much 
employed in the literature. 
An advantage of the real-time formalism is the possibility of
identifying an enhanced tunneling rate at a finite time of
dynamical evolution of the tunneling \cite{my 00-3}.

Our formalism is based on separation of a subsystem from thermal
environment, and integrating out the environment degrees of freedom.
This method is  best suited to a (by itself)
non-equilibrium system which is immersed in a larger thermal equilibrium
state.
Detailed properties of the environment and its interaction form with
the subsystem are expected to be insensitive to the behavior of the
subsystem in question.
We use the standard model of environment consisting of an infinite number of
harmonic oscillators \cite{caldeira-leggett 83}, \cite{feynman-vernon}.
In this picture dissipation seen in the behavior of the subsystem 
is due to our ignorance of the huge environment degrees of freedom.
Although the subsystem itself is an open system having an interaction with
environment, the entire system including the environment obeys 
quantum mechanical laws in the closed system.

We work out consequences of the real-time formalism for
a simple subsystem in one dimension.
The most interesting result that comes out in this study is
the resonance enhanced tunneling \cite{my 00-3}. The enhancement occurs at
a time scale of order $1/\eta$, with $\eta$ the friction
coefficient. The enhanced tunneling probability goes with $\eta$ 
like $\eta^{- 1.5}$ according to our numerical analysis in the present work. 
We offer an interpretation of the resonance enhancement due
to an energy flow into the system from environment.
Furthermore, the enhanced tunneling rate is related to the phenomenon
of the parametric resonance. We call this enhanced tunneling as 
environment-driven excited tunneling (EET). 
The popular Euclidean approach cannot deal with the finite time
behavior of tunneling, hence misses the resonance enhancement
at finite times.

Our method, with a certain extension, should be applicable
to field theory models. In future we hope to investigate applications
to cosmology, in which one may deal with the electroweak first order
phase transition relevant to baryogenesis \cite{ew-bgeneration-review}.
An important question here is the time scale of tunneling.
If the usual Hubble time scale that
gives rise to the potential (or more properly the free energy) change,
hence regarded in the conventional picture 
as the termination of the first order phase transition,
is replaced by a shorter tunneling time scale of order $1/\eta$, 
the conventional picture of the first
order phase transition may drastically be changed.
For instance, bubbles of the true vacuum may be formed by the
resonance enhanced tunneling prior to the potential change. 
Since this process is stochastic,
it may enhance the out-of-equilibrium condition necessary for
baryogenesis \cite{sakharov}, \cite{my}.
The bubble formation via EET may also
resurrect the old scenario of GUT phase transition \cite{old inflation}
which was once discarded due to the graceful exit problem
\cite{problem of old inflation}.

The rest of this paper is organized as follows.
In Section \lromn 2 the model and the basic tool of analysis
is explained.
In Section \lromn 3 the semiclassical approximation is introduced
and the result of the crucial simplification, the linear approximation
of the environment variable, is worked out.
This approximation is used while the particle in one dimensional potential
is in the metastable region, to determine how the particle is excited
in the potential well due to the environment interaction.
In Section \lromn 4 detailed numerical analysis in the metastable region
is presented, taking a concrete example of the potential.
It is shown that excitation to higher levels proceeds via
the parametric resonance effect in the linear approximation.
In Section \lromn 5 the semiclassical formula in preceding sections
is combined with the quantum mechanical barrier penetration factor
to derive the tunneling rate in thermal medium. Numerical analysis
is then presented, to quantitatively demonstrate the effect of EET.
In the final section application to cosmological baryogenesis
is briefly discussed.
In Appendix we give the friction coefficient along with relevant
formulas in the standard model of particle physics, which
may be applicable to discussion of the electroweak baryogenesis
in the standard model.

%%%%%%%%%%%%%%%%%%%%%%%%%%%%%%%%%%%%%%%%%%%%%%%%%%%%%%%%%%%%%%%%%%%%%%%%%%%

\vspace{1cm}
\lromn 2 \hspace{0.2cm} {\bf Model and basic tool}
\vspace{0.5cm} 

\begin{figure}[t]
  \begin{center}
    \includegraphics[height = 5cm,clip]{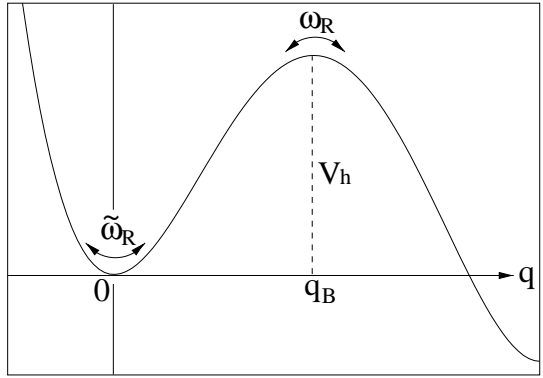}
  \end{center}
  \caption{\small
  A typical potential for tunneling in one dimensional system.
  Curvature parameters, $\tilde{\omega}_R$ at the bottom of the potential,
  $\omega_R $ at the top of the potential, along with the barrier height $V_h$
  and the barrier width, characterize a global structure of the potential.
  }
  \label{V}
\end{figure}

We consider as the first step of
our investigation the simplest, yet the most basic problem of this kind,
one dimensional system described by a potential $V(q)$.
The potential $V(q)$, as illustrated in Fig.\ref{V}, 
is assumed to have some local minimum at
$q = 0$ with $V(0) = 0$, 
which is separated at the barrier top, $q = q_{B} \,(\,> 0)$,
from a global minimum, which we take in the present work
to be at $q = \infty$ such that there is no reflection
from a wall at the far right.
This system is put in thermal medium which acts as an environment.
It has infinitely many, continuously
distributed harmonic oscillators given by their coordinates $Q(\omega )$.
Its coupling to the tunneling system is given by a Hamiltonian
\cite{caldeira-leggett 83}, \cite{feynman-vernon},
\begin{equation}
q\,\int_{\omega _{c}}^{\infty }\,d\omega \,c(\omega )Q(\omega )
\,.
\end{equation}
The frequency dependent coupling strength is 
$c(\omega )$ and $\omega _{c}$ is some threshold frequency.
One may imagine a generalized case in which the subsystem variable
$q$ is the order parameter for the first order phase transition
in cosmology,
the homogeneous Higgs field, and the environment oscillator
$Q(\omega )$ is a collection of various forms of matter fields
coupled to the Higgs field.

The basic dynamical equation in our problem is then
\begin{eqnarray}
&&
\frac{d^{2}q}{dt^{2}} + \frac{dV}{dq} =
-\,\int_{\omega_c}^{\infty}\,d\omega \,c(\omega )Q(\omega ) \,, 
\label{eq of motion 1}
\\ &&
\frac{d^{2}Q(\omega )}{dt^{2}} + \omega ^{2}\,Q(\omega )
= -\,c(\omega )\,q
 \,.
\label{eq of motion 2}
\end{eqnarray}
One may eliminate the environment variable $Q(\omega )$ from
the second equation above, to get
the Langevin equation \cite{ford-lewis-oconnell},
\begin{eqnarray}
&&
\frac{d^{2}q}{dt^{2}} + \frac{d V}{d q} +
2\,\int_{0}^{t}\,ds\,\alpha _{I}(t - s)q(s) = F_{Q}(t) 
\,.
\label{langevin eq}
\end{eqnarray}
The kernel function $\alpha _{I}$ here 
is given by
\begin{equation}
\alpha _{I}(t) = -\,\int_{\omega _{c}}^{\infty }\,d\omega \,
r(\omega )\sin (\omega t) 
 \,,
\end{equation}
with 
\begin{equation}
r(\omega ) = \frac{c^{2}(\omega )}{2\omega}
\,.
\end{equation}
By definition the spectral function $r(\omega )$ is positive definite.
The random force term $F_{Q}(t) $ in eq.(\ref{langevin eq}) 
is linear in initial values 
of environment variables, $Q_{i}(\omega )$ and 
$P_{i}(\omega) = \dot{Q}_{i}(\omega )$ at
$t = 0$;
\begin{eqnarray}
&&
F_{Q}(t) = -\,
\int_{\omega _{c}}^{\infty }\,d\omega \,c(\omega )\,
\left( \,Q_{i}(\omega ) \cos (\omega t) + P_{i}(\omega)
\frac{\sin (\omega t)}{\omega } \,\right)
\,.
\end{eqnarray}
By taking the thermal bath of temperature $T = 1/\beta $
given by the density matrix,
\begin{eqnarray}
\rho _{\beta }(Q \,, Q') &=& \left( \frac{\omega }{\pi \,\coth (\beta \omega 
/2)}\right)^{1/2}\,
\nonumber \\ &&
\cdot \exp \left[ \,-\,\frac{\omega }{2 \sinh (\beta \omega )}
\,\left( \,(Q^{2} + Q'\,^{2})\,\cosh (\beta \omega ) - 2 Q Q'\,\right)\,\right]
\,,
\label{thermal density matrix} 
\end{eqnarray}
for each environment oscillator,
the following correlation formula in thermal medium is obtained;
\begin{eqnarray}
&&
\hspace*{-0.5cm}
\langle \{ F_{Q}(\tau )\,,F_{Q}(s) \}_{+} \rangle_{{\rm env}}
= \int_{\omega _{c}}^{\infty }\,d\omega \,r(\omega )
\cos \omega (\tau - s)\,\coth (\frac{\beta \omega }{2}) 
\equiv \alpha _{R}(\tau - s)
\,.
\end{eqnarray}
The combination of the two real 
$\alpha$'s, $\alpha _{R}(t) + i\,\alpha _{I}(t)$,
is  the well-known real-time thermal Green's function 
\cite{Fetter-Walecka} summed over $Q(\omega)-$oscillators, being added with
the weight $c^2(\omega)$.

An often used simplification is the local, Ohmic approximation
taking 
\begin{equation}
r(\omega) = \frac{\eta\,\omega}{\pi} \,,
\label{ohmic model}
\end{equation}
with $\omega_c = 0$,
which amounts to \cite{caldeira-leggett 83-0}
\begin{equation}
\alpha _{I}(\tau ) = \delta \omega ^{2}\delta (\tau )
+ \eta\, \delta '(\tau ) \,.
\end{equation}
This gives the local version of the Langevin equation,
\begin{equation}
\frac{d^{2}q}{dt^{2}} + \frac{d V}{d q} + 
\delta \omega ^{2}\,q + \eta\,\frac{dq}{dt} = F_Q \,.
\end{equation}
The parameter $\delta \omega ^{2}$ is interpreted as
a potential renormalization or a mass renormalization in the
field theory analogy, since by changing the bare frequency parameter
to the renormalized $\omega _{R}^{2}$ the term
$\delta \omega ^{2}\,q$ is cancelled by a counter term
in the potential.
On the other hand, $\eta $ is the Ohmic friction coefficient.
This local approximation breaks down both at early and at late
times \cite{difficulty of ohm}, 
but it is useful in other time regions.

The value of the coefficient of quadratic term of the potential is
shifted by environment interaction. We discuss the relation between the
renormalized curvature $\omega_R$ ($\tilde{\omega}_R$) at the barrier
top (the potential bottom) and the physical curvature $\omega_B$
($\omega_p$) here. These physical curvatures $\omega_B, \omega_p$ are
frequently used in the following discussion. We first illustrate this
for the curvature at the barrier top. In the vicinity of the barrier
top, the renormalized potential is given by $V(q) \simeq -\omega_R^2(q -
q_B)^2/2 + V_h + {\rm counter ~terms}~~~(\omega_R > 0 {\rm
~by~definition})$. The physical curvature $\omega_B$ (the tachyon mass
in the field theory analogy, which is defined by a positive value)
shifted by the environment is determined as a pole position of the
Green's function $\langle 0|T[q(t)q(0)]|0\rangle$ at the barrier top,
thus as a solution for the isolated pole of the following equation; 
\begin{eqnarray}
&&
\omega _{B}^{2} - \omega _{R}^{2} + \int_{\omega _{c}}^{\infty }\,
d\omega \,\frac{2\omega_B^2\,r(\omega )}
{\omega(\omega _{B}^{2} + \omega^{2})} = 0
\,.
\label{exact pole location} 
\end{eqnarray}

In many past works, the renormalized curvature $\omega _{R}$ was used
as a reference curvature to evaluate the environment effect. Due to the 
positivity of $r(\omega )$, one has $\omega _{B}^{2} < \omega _{R}^{2}$
in general. A consequence of this is that the potential barrier 
is further suppressed by using $\omega_B$ instead of taking 
the renormalized curvature $\omega _{R}$ as the reference point.
It is instructive to give an example.
In the Ohmic model of eq.(\ref{ohmic model})
the explicit form of the physical curvature $\omega _{B}$
is given by
\begin{equation}
\omega _{B} = \sqrt{\omega _{R}^{2} + \frac{\eta ^{2}}{4}} 
- \frac{\eta }{2} \,.
\label{ohmic relation for b and r} 
\end{equation}
The relation between the renormalized and the physical curvature at the
local minimum of the potential is obtained in the same way, 
to give in the Ohmic model
\begin{equation}
\omega _{p} = \sqrt{\,\tilde{\omega} _{R}^{2} - \frac{\eta ^{2}}{4}\,}
\,.
\label{pole position}
\end{equation}
In subsequent discussion we shall specify which parameter we use.

We now turn to description of the macroscopic system determined by
dynamical coordinates $q \,, Q(\omega)$.
Any mixture of quantum pure states, namely a quantum statistical state, 
is described by the density matrix,
\begin{equation}
\rho = \sum_{n = 0}^{\infty} \,w_{n}\,|n\rangle \langle n |\,, 
\label{def density matrix} 
\end{equation}
where $|n\rangle$ is a linearly independent 
state vector for a pure quantum state and
$0 \leq w_{n} \leq 1$.
As a special case one may deal with a pure state taking
$w_{n} = 0$ for $n \neq 0$.
The density matrix that describes the entire quantum mechanical
system obeys the equation of motion;
\begin{equation}
i\hbar \frac{\partial \rho }{\partial t} =
\left[ \,H _{{\rm tot}} \,, \: \rho \,\right]
\,, 
\label{liouville eq} 
\end{equation}
where $H _{{\rm tot}}$ is the total Hamiltonian of the entire
system.
For the time being, we explicitly write the Planck constant $\hbar$.
In the configuration space this density matrix is given by its
matrix elements,
\begin{equation}
\rho (q \,, q' \,; \; Q(\omega ) \,, Q'(\omega ) )
= \langle q \,, Q(\omega )|\,\rho \,|q' \,, Q'(\omega ) \rangle
\,.
\end{equation}
Its Fourier transform with respect to relative coordinates,
\( \:
q - q' \,, \, Q(\omega ) - Q'(\omega )
\,, 
\: \)
is called the Wigner function and is denoted by $f_{W}$;
\begin{eqnarray}
&&
f_{W}(q \,, p \,; \; Q(\omega )\,, P(\omega ) )
= \int_{-\infty }^{\infty }\,d\xi \,\prod_{\omega}\,dX(\omega )\,
\exp [-\,i\xi p - i\int\,d\omega \,X(\omega )P(\omega )]\,
\nonumber \\ &&
\hspace*{0.5cm}
\times \,
\rho (q + \xi /2\,, q - \xi /2 \,; \; Q(\omega ) +
X(\omega )/2\,, Q(\omega ) - X(\omega )/2 ) \,.
\end{eqnarray}
It is easy to show from the master equation
(\ref{liouville eq}) that this quantity obeys ;
\begin{eqnarray}
&&
\frac{\partial f_{W}}{\partial t} =
-\,p\,\frac{\partial f_{W}}{\partial q}
+ \frac{1}{i\hbar }\,
\left\{ \,
V\left( q + \frac{i\hbar }{2}\frac{\partial }{\partial p}\right) 
- V\left( q - \frac{i\hbar }{2}\frac{\partial }{\partial p}\right) 
\,\right\}
f_{W}
\label{master eq for wigner} 
\\
&&- \int\,d\omega \,
\left( \,
P(\omega )\,\frac{\partial f_{W}}{\partial Q(\omega )}
+ \omega ^{2}\,Q(\omega )\,
\frac{\partial f_{W}}{\partial P(\omega )}
+\, c(\omega )\,(q\frac{\partial }{\partial P(\omega )}
+ Q(\omega )\frac{\partial }{\partial p})\,f_{W}
\,\right)
\nonumber\,. 
\end{eqnarray}

Reduction to the subsystem by integrating out
environment variables $Q(\omega)$ is an essential part for
our physical understanding of tunneling in thermal medium.
We define the reduced density matrix of the tunneling system by
\begin{eqnarray}
&&
\rho^{(R)}(q \,, q') =
\int\, dQ(\omega )\,dQ'(\omega)\,
\,\delta(Q(\omega )- Q'(\omega))
\cdot 
\rho (q \,, q' \,; \; Q(\omega ) \,, Q'(\omega ) )
\,.
\end{eqnarray}
The fact that there is a delta function $\delta (Q(\omega )- Q'(\omega))$ 
for the environment variable
reflects our understanding that one does not observe the environment.
A similar reduction is made to the Wigner function, to get
the reduced Wigner function $f_W^{(R)}(q \,, p)$;
\begin{eqnarray}
  f_W^{(R)}(q, p) 
  = \int dP(\omega)~dQ(\omega) f_W(q,p;Q(\omega), P(\omega) )
\,.
\end{eqnarray}

The reduction to the subsystem here is in spirit opposite to
that common in the effective field theory approach;
integration over heavy degrees of freedom is performed.
What is integrated out in the present case is the majority of
constituents which is taken invariant under frequent thermalizing
interaction.

%%%%%%%%%%%%%%%%%%%%%%%%%%%%%%%%%%%%%%%%%%%%%%%%%%%%%%%%%%%%%%%%%%%%%%%%%%%

\vspace{1cm}
\lromn 3 \hspace{0.2cm} {\bf Basic idea and semiclassical Wigner function}
\vspace{0.5cm}

In considering our problem of tunneling in thermal medium
we assume that effects of thermal environment
are most important in the time period when
 the particle in one dimensional system
remains in the metastable potential well.
Thus, we anticipate that at tunneling
the simple quantum mechanical formula may directly be used.
We thus separate the problem into two parts;
first to treat the potential region left to the barrier top,
$q < q_B$, by semiclassically integrating out environment variables
and then by projecting the derived reduced density
matrix onto the energy eigenstate of the subsystem,
which makes it possible to convolute 
the usual barrier penetration factor used in quantum mechanics.
In the following two sections we do this and derive a formula
for the tunneling rate in thermal medium.

In the semiclassical $\hbar \rightarrow 0$ limit we have
in eq.(\ref{master eq for wigner})
\begin{eqnarray}
\frac{1}{i\hbar }\,
\left\{ \,
V\left( q + \frac{i\hbar }{2}\frac{\partial }{\partial p}\right) 
- V\left( q - \frac{i\hbar }{2}\frac{\partial }{\partial p}\right) 
\,\right\}
\,\rightarrow\,
\frac{d V}{d q}\,\frac{\partial }{\partial p}
\,.
\end{eqnarray}
The resulting equation for the Wigner function, 
\begin{eqnarray}
&&
\frac{\partial f_{W}}{\partial t} =
-\,p\,\frac{\partial f_{W}}{\partial q}
+ \frac{d V}{d q}\,\frac{\partial f_{W}}{\partial p}
\label{semiclassical master eq for wigner} 
\\
&&- \int\,d\omega \,
\left( \,
P(\omega )\,\frac{\partial f_{W}}{\partial Q(\omega )}
+ \omega ^{2}\,Q(\omega )\,
\frac{\partial f_{W}}{\partial P(\omega )}
+\, c(\omega )\,(q\frac{\partial }{\partial P(\omega )}
+ Q(\omega )\frac{\partial }{\partial p})\,f_{W}
\,\right)
\,. 
\nonumber
\end{eqnarray}
is identical to the classical Liouville equation familiar in
classical statistical mechanics. 
It has an obvious, formal solution;
\begin{eqnarray}
&&
f_{W}(q\,, p\,, Q\,, P) = 
\int\,dq_{i}dp_{i}\,\int\,dQ_{i}dP_{i}\,
f_{W}^{(i)}(q_{i}\,, p_{i}\,, Q_{i}\,, P_{i})\,
\nonumber \\ &&
\hspace*{1cm} 
\cdot 
\delta \left( q - q_{{\rm cl}}\right)\,\delta \left( p - p_{{\rm cl}}\right)
\,\delta \left( Q - Q_{{\rm cl}}\right)\,\delta \left( P 
- P_{{\rm cl}}\right)
 \,,
\end{eqnarray}
where 
\( \:
q_{{\rm cl}}
\,, Q_{{\rm cl}}(\omega) 
\: \)
etc. are the solution of (\ref{eq of motion 1}), (\ref{eq of motion 2}) 
taken as the classical equation with the specified initial condition
written in terms of $q_i \,, p_i \,, Q_i \,, P_i $.

We consider the circumstance under which the tunneling system
is initially in a state uncorrelated to the rest of environment.
Thus we take the form of the initial Wigner function of 
a factorized form,
\begin{eqnarray}
  f_{W}^{(i)}(q_{i}\,, p_{i}\,, Q_{i}\,, P_{i})\,
  =
  f_{W\,, q}^{(i)}(q_{i}\,, p_{i})
  \times
  f_{W\,, Q}^{(i)}(Q_{i}\,, P_{i})~,
\end{eqnarray}
to get the reduced Wigner function after the environment
$Q(\omega) \,, P(\omega)$ integration,
\begin{eqnarray}
&&
f_{W}^{(R)}(q \,, p\,;t) = 
\int\,dq_{i}dp_{i}\,
f_{W\,, q}^{(i)}(q_{i}\,, p_{i})\,K(q \,, p\,, q_{i}\,, p_{i}\,;t)
\,,
\label{integral transform} 
\\ && 
K(q \,, p\,, q_{i}\,, p_{i}\,;t) = \int\,dQ_{i}dP_{i}\,
f_{W\,, Q}^{(i)}(Q_{i}\,, P_{i})\,
\delta \left( q - q_{{\rm cl}}\right)\,\delta \left( p - p_{{\rm cl}}\right)
 \,.
\end{eqnarray}
The quantity $q_{{\rm cl}}$ is now a solution of the Langevin equation 
(\ref{langevin eq}),
regarded as a function of initial values;
$q_{{\rm cl}}(q_{i}\,, p_{i}\,, Q_{i}\,, P_{i}\, ; t)$.
The problem of great interest is how further one can practically simplify
the kernel function $K$ introduced here.

In many situations one is interested in the tunneling probability when
the environment temperature is low enough.
At low temperatures of
\( \:
T \ll 
\: \)
a typical frequency or curvature scale 
$\omega _{s}$ of the potential $V$,
one has
\begin{equation}
\omega _{s}\,\sqrt{\,\overline{Q_{i}^{2}(\omega )}\,} \,, 
\sqrt{\,\overline{P_{i}^{2}(\omega )}\,} = O[\sqrt{T}]
\ll \sqrt{\omega _{s}}
 \,.
\end{equation}
In the electroweak and the GUT phase transition this
corresponds to $T \ll m_{H}$(Higgs mass).
Expansion of $q_{{\rm cl}}$ in terms of 
\( \:
Q_{i}(\omega ) \,, P_{i}(\omega )
\: \)
and its truncation to linear terms is then justified.
Thus, we use
\begin{eqnarray}
&&
\delta \left( \,q - q_{{\rm cl}}\,\right)
= \int\,\frac{d\lambda _{q}}{2\pi }\,
\exp \left[ \,i\,\lambda _{q}\,\left( \,q - q_{{\rm cl}}\,\right)\,\right]
\nonumber \\ &&
\hspace*{-1cm}
\approx \int\,\frac{d\lambda _{q}}{2\pi }\,
\exp \left[ \,i\,\lambda _{q}\,
\left( \,q - q_{{\rm cl}}^{(0)} -
\int\,d\omega \,
\left\{\,Q_{i}(\omega )q_{{\rm cl}}^{(Q)}(\omega )
+ P_{i}(\omega )\,q_{{\rm cl}}^{(P)}(\omega )\,
\right\}\,\right)\,\right]
 \,,
\label{expansion of cl sol}
\end{eqnarray}
valid to the first order of $Q_{i}(\omega) \,, P_{i}(\omega)$.
The zero-th order term is $q_{{\rm cl}}^{(0)}$ obeying
\begin{eqnarray}
\frac{d^{2}q_{{\rm cl}}^{(0)}}{dt^{2}} +
\left( \frac{d V}{dq }\right)_{q_{{\rm cl}}^{(0)}}
+ 2\,\int_{0}^{t}\,ds\,\alpha _{I}(t - s)q_{{\rm cl}}^{(0)}(s) = 0
 \,,
\label{homo classical eq}
\end{eqnarray}
while the coefficient functions of the first order terms are
$q_{{\rm cl}}^{(Q)}(\omega ) \,, q_{{\rm cl}}^{(P)}(\omega )$
obeying equations later shown, eq.(\ref{fluctuation eq}).
A similar expansion for
\( \:
\delta \left( \,p - p_{{\rm cl}}\,\right)
\: \)
using
\( \:
p_{{\rm cl}}^{(0)} \,, p_{{\rm cl}}^{(Q)}(\omega ) \,,
p_{{\rm cl}}^{(P)}(\omega )
 \,,
\: \)
also holds.

An alternative justification of the expansion (\ref{expansion of cl sol})
is to keep $O[c(\omega)]$ terms consistently in the exponent
since both $q_{{\rm cl}}^{(Q)}(\omega )$ and $q_{{\rm cl}}^{(P)}(\omega )$
are of this order.
Thus, both in the low temperature and in the weak coupling case
our subsequent analysis follows.

With (\ref{expansion of cl sol}) one can proceed further.
Gaussian integral for the variables
$Q_i(\omega)$, $P_i(\omega)$,
first, and then for
\( \:
 \lambda _{q} \,, \lambda _{p}
\: \)
can be done explicitly with eq.(\ref{expansion of cl sol}), 
when one takes the thermal, hence Gaussian density matrix
for the initial environment variable:
\begin{eqnarray}
&&
f_{W\,, Q}^{(i)}(Q_{i}\,, P_{i})
=
\frac{1}{\pi}\tanh \frac{\beta \omega}{2}
\exp
\left[
  -\tanh \frac{\beta \omega}{2}
  \left(
    \omega Q_{i}^2 + \frac{P_{i}^2}{\omega}
  \right)
\right]
\,,
\end{eqnarray}
The result of this Gaussian integral leads to
an integral transform \cite{my 00-2} of the Wigner function, 
$f_{W}^{(i)} $ (initial) $\rightarrow f_{W}^{(R)}$ (final), 
using the following form of the kernel function;
\begin{eqnarray}
&&
\hspace*{-1cm}
K(q \,, p\,, q_{i}\,, p_{i}\,;t) = 
\frac{\sqrt{{\rm det}\; {\cal J}\,}}{2\pi}
\exp \left[ \,-\,\frac{1}{2}
(q - q_{{\rm cl}}^{(0)}\,, \, p - p_{{\rm cl}}^{(0)})
\,{\cal J}\,
\left( \begin{array}{c}
q - q_{{\rm cl}}^{(0)}  \\
p - p_{{\rm cl}}^{(0)}
\end{array}
\right)
\,\right]
 \,,
 \label{kernel function} 
\end{eqnarray}
where the matrix elements of 
\( \:
{\cal J}^{-1}_{i j} = I_{ij}
\: \)
are given by 
\begin{eqnarray}
I_{11}(t) &=& \frac{1}{2}\,\int_{\omega _{c}}^{\infty }\,d\omega \,
\coth \frac{\beta \omega }{2}\,
\frac{1}{\omega } \,|z(\omega \,, t)|^2
\,, 
\label{fluctuation 11}
\\
I_{22}(t) &=& \frac{1}{2}\,\int_{\omega _{c}}^{\infty }\,d\omega \,
\coth \frac{\beta \omega }{2}\,
\frac{1}{\omega } \,
|\dot{z}(\omega \,, t)|^2
\,, 
\\
\hspace*{-1cm}
I_{12}(t)  &=& \frac{1}{2}\,\int_{\omega _{c}}^{\infty }\,d\omega \,
\coth \frac{\beta \omega }{2}\,
\frac{1}{\omega } \,\Re\left( z(\omega \,, t)\dot{z}^*(\omega \,, t)\right)
= \frac{\dot{I}_{11}}{2}
\,,
\label{fluctuation 12}
\end{eqnarray}
where
\( \:
z(\omega \,, t) = 
q_{{\rm cl}}^{(Q)}(\omega \,, t) + i\,
\omega \,q_{{\rm cl}}^{(P)}(\omega \,, t)\,,
\: \)
and 
\( \:
\dot{z}(\omega \,, t) = 
p_{{\rm cl}}^{(Q)}(\omega \,, t) +
i\omega \,p_{{\rm cl}}^{(P)}(\omega \,, t)
\,.
\: \)

Quantities that appear in the integral transform are determined by solving
differential equations; 
the homogeneous Langevin equation for $q_{{\rm cl}}^{(0)}$, 
eq.(\ref{homo classical eq})
and an inhomogeneous linear equation for $z(\omega \,, t)$ and
$\dot{z}(\omega \,, t)$,
\begin{eqnarray}
&&
\frac{d^{2}z(\omega \,, t)}{dt^{2}} +
\left( \frac{d^{2} V}{dq^{2} }\right)_{q_{{\rm cl}}^{(0)}}\,z(\omega \,, t)
+ 2\,\int_{0}^{t}\,ds\,\alpha _{I}(t - s)z(\omega \,, s) = 
-\,c(\omega )e^{i\omega t} \,.
\label{fluctuation eq}
\end{eqnarray}
A similar equation as for $z(\omega \,, t)$ holds for
\( \:
\dot{z}(\omega \,, t)  \,.
\: \)
The initial condition is 
\begin{eqnarray}
q_{{\rm cl}}^{(0)}(t = 0) = q_{i} \,, \hspace{0.5cm} 
p_{{\rm cl}}^{(0)}(t = 0) = p_{i} \,, \hspace{0.5cm} 
z(\omega \,, t = 0) = 0  \,, \hspace{0.5cm}
\dot{z}(\omega \,, t = 0) = 0 \,.
\end{eqnarray}

The physical picture underlying the formula for the integral
transform, eq.(\ref{integral transform}) along with 
(\ref{kernel function}), should be evident;
the probability at a phase space point $(q\,,p)$ is dominated 
by the semiclassical trajectory $q_{{\rm cl}}^{(0)}$
(environment effect of dissipation being included in its determination
by the third term of eq.(\ref{homo classical eq}))
reaching $(q\,,p)$ from an initial point $(q_i \,, p_i)$ 
whose contribution is weighed by the probability 
$f^{(i)}_W$ initially given.
The contributing trajectory is broadened by the environment
interaction with the width factor $\sqrt{I_{ij}}$.
The quantity $I_{11}$ given by (\ref{fluctuation 11}),
for instance, is equal to
\( \:
\overline{(q - q_{{\rm cl}}^{(0)})^{2}}
\,;
\: \)
an environment driven fluctuation under the stochastic
force $F_{Q}(t)$.

We note that in the exactly solvable model of inverted
harmonic oscillator potential the identical form of
the integral transform eq.(\ref{kernel function}) 
was derived \cite{my 00-1} without resort to
the semiclassical approximation.
Explicit form of the classical and the fluctuation
functions ($q_{{\rm cl}}^{(0)}$ and $z(\omega \,, t)$) is given there,
which is valid beyond the Ohmic approximation.

It would be instructive to write solutions in the harmonic
approximation taking the frequency at the local potential minimum
with
\( \:
\omega_p = \sqrt{\tilde{\omega} _{R}^{2} - \frac{\eta ^{2}}{4}} \,,
\: \)
eq.(\ref{pole position}),
the potential is given by
$V(q) = \frac{1}{2}\,\omega_p ^2\, q^2$.
Then the zero-th order term $q^{(0)}_{\rm cl}(t)$ 
and $z(\omega,t)$ are obtained analytically,
and given by
\begin{eqnarray}
&&
q_{{\rm cl}}^{(0)} = 
u(t)\,q_{i} + v(t)\,p_{i} \,, 
\\ &&
u(t) = 
(\cos \omega_p t + \frac{\eta }{2\omega_p}\sin \omega_p t)\,e^{-\eta t/2}
\,, 
\hspace{0.5cm} 
v(t) = \frac{\sin \omega_p t}{\omega_p}\,e^{-\eta t/2}
\,, 
\label{def of u,v} 
\\ && 
z(\omega \,, t) =
\frac{c(\omega )}{\omega ^{2} - \tilde{\omega} _{R}^{2} - i\omega \eta }\,
\left( \,e^{i\omega t} - u(t) - i\omega v(t) \,\right)
 \,.
\end{eqnarray}
We used the local Ohmic approximation such as
\begin{equation}
2\,\int_{0}^{t}\,ds\,\alpha _{I}(t - s)z(\omega \,, s) 
\;\rightarrow\; \eta\,\dot{z}(\omega \,, t) \,.
\end{equation}
The function $z(\omega \,, t)$ has a Breit-Wigner form for a small
$\eta $.
In particular, the infinite time ($t \rightarrow \infty $)
limit of the fluctuation has
\begin{equation}
I_{11}(\infty ) =
\int_{\omega _{c}}^{\infty }\,d\omega \,\coth \frac{\beta \omega }{2}\,
\frac{r(\omega )}{(\omega ^{2} - \tilde{\omega}_{R}^{2})^{2} + \eta ^{2}\omega ^{2}} 
\,.
\label{infinite time limit of fluc}
\end{equation}
The friction $\eta$ appears as the resonance width in this formula.
The real part of the pole position in the frequency integral such as
(\ref{infinite time limit of fluc}),
$\omega_p \equiv \Re $ (pole position), is $\omega_p = \sqrt{\tilde{\omega} _{R}^{2} - \frac{\eta ^{2}}{4}}
\,.$

The harmonic approximation is excellent at late times, but
at early and intermediate times nonlinear resonance effects
are important as discussed in \cite{my 00-3}.
The nonlinear effect will be taken into account in discussion of the
finite time behavior of the tunneling rate.

A convenient form of the fluctuation formula is derived
by separating the zero-temperature result along with
$\coth \beta \omega /2 = 1 + 2/(e^{\beta \omega } - 1)$.
We denote the temperature dependent part of the fluctuation
by $I_{ij}^{(T)}$.
The zero-temperature part $I_{ij}^{(0)}$ must be renormalized.
After performing renormalization the zero-temperature part
is approximated by result of the harmonic approximation.
The renormalized fluctuation is then well approximated using the
physical curvature, $\omega _{p}$, eq.(\ref{pole position}).
We shall use the same notation $I_{ij}$ for the renormalized
fluctuation in the rest of our paper.

%%%%%%%%%%%%%%%%%%%%%%%%%%%%%%%%%%%%%%%%%%%%%%%%%%%%%%%%%%%%%%%%%%%%%%%%%%%

\vspace{1cm}
\lromn 4 \hspace{0.2cm} {\bf Numerical analysis in the metastable
region; relevance of parametric resonance}
\vspace{0.5cm}

We point out a relation of the semiclassical Wigner function in the
preceding section to the parametric resonance,
as briefly mentioned in \cite{my 00-3}.
Presence of anharmonic terms is inevitable in any realistic
tunneling potential.
This introduces a non-trivial oscillating term in the coefficient
function 
$\left( \frac{d^{2} V}{dq^{2} }\right)_{q_{{\rm cl}}^{(0)}}$
of the fluctuation equation, (\ref{fluctuation eq}),
thus giving a differential equation akin to the Mathieu equation.
Depending on whether or not the relevant parameters fall into
the instability band of this Mathiew-type equation, the fluctuation
may increase indefinitely in the zero friction limit
\cite{Landau mechanics}.
What happens is more subtle; the parameters fall right on
the boundary of stability and instability bands.
We shall show this by taking an example of the tunneling potential.

\begin{figure}[t]
  \begin{center}
    \includegraphics[height = 5cm,clip]{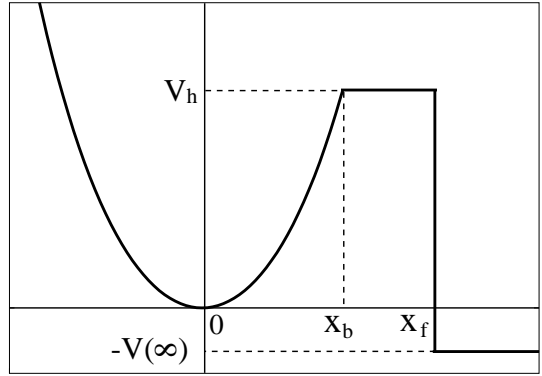}
  \end{center}
  \caption{\small 
  The tunneling potential used in numerical calculation.}
  \label{concV}
\end{figure}
The potential we take (illustrated in Fig.\ref{concV}) is given by
\begin{eqnarray}
V(x) &=& \frac{\tilde{\omega}_R^2}{2}\,x^2 + g\,\tilde{\omega}_R^3\, x^4 
\,,         
\hspace{0.5cm}
(x < x_b)
\nonumber
\\ 
&=&    V_h    \,,          \hspace{0.5cm}
                           (x_b < x < x_f)
\,,
\hspace{0.5cm}
=  -\,V(\infty)      \,,  \hspace{0.5cm}                           (x_f < x)
\,.
\label{potential}
\end{eqnarray}
The two parameters $V_h$ and $x_b$ are related since we took 
a particular quartic potential in the metastable well.

The fluctuation $z(\omega,t)$ in the Ohmic approximation follows the equation;
\begin{equation}
 \ddot{z}(\omega,t) + V''( q_{{\rm cl}}^{(0)} )\,z(\omega,t) + 
\eta\,\dot{z}(\omega,t) = -\,c(\omega)\,e^{i\omega t}
\,.
\label{inhomogeneous fluc eq}
\end{equation}
The potential derivative here is
\begin{equation}
V''( q_{{\rm cl}}^{(0)} ) = \tilde{\omega}_R^2 + 12g\,\tilde{\omega}_R^3\, 
( q_{{\rm cl}}^{(0)} )^2
\,.
\end{equation}
Taking the zero friction limit $\eta = 0$ gives the following related
homogeneous equation,
\begin{equation}
\ddot{z}(\omega,t) + [h + 2\theta\, ( q_{{\rm cl}}^{(0)} )^2] z(\omega,t) = 0
\,,
\label{homogeneous eq in zero friction}
\end{equation}
where
\begin{equation}
h = \tilde{\omega}_R^2 \,, \hspace{0.5cm} \theta = 6g\,\tilde{\omega}_R^3
\,.
\label{model parameter region}
\end{equation}
The equation for $q_{{\rm cl}}^{(0)}$ is as follows:
\begin{eqnarray}
  \ddot{q}_{\rm cl}^{(0)}
  +
  V'(q_{\rm cl}^{(0)})
  =
  0
\end{eqnarray}
This gives a periodic solution for $q_{{\rm cl}}^{(0)}$.
A notable feature of the homogeneous equation in the $\eta = 0$ limit 
(\ref{homogeneous eq in zero friction}) is that $q_{{\rm cl}}^{(0)}$ 
appearing in the coefficient function is periodic, hence
it gives rise to solutions either of the Bloch wave type or of the parametric
resonance type \cite{Landau mechanics}.
Despite of the definite relation, eq.(\ref{model parameter region}), 
we regard $(h, \theta)$ as free parameters 
to throughly investigate the stable-unstable band structures,
because that would more definitely clarify relation between
the band structure and the relevant parameter region of this model.

\begin{figure}[t]
  \begin{center}
    \includegraphics[height = 4.3cm,clip]{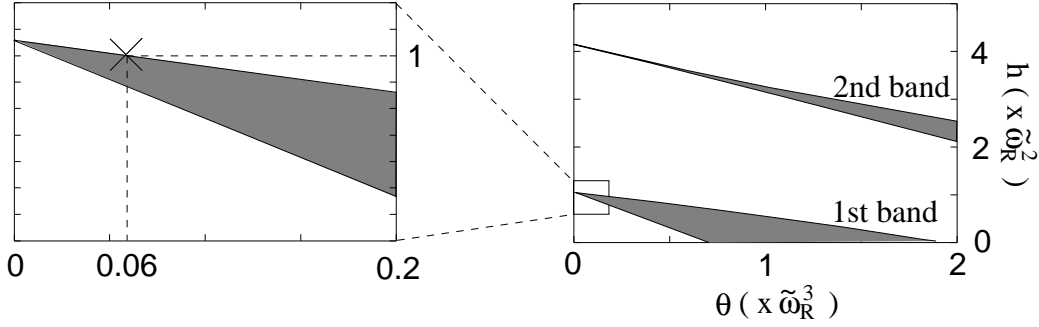}
  \end{center}
  \caption{\small
  Band structure for relevant fluctuation.
  Shaded areas correspond to the instability band.
  1st band is enlarged separately (the left figure), where the crossed point 
  corresponds to our model relation, eq.(\ref{model parameter region}).
  }
  \label{h_th}
\end{figure}

For simplicity we analyzed the problem assuming
\begin{equation}
g = 0.01 \,, \hspace{0.5cm}
V_h = 5\omega_p \,, \hspace{0.5cm}
\omega_p = \sqrt{\,\tilde{\omega}_R^2 - \eta^2/4\,}
\,.
\end{equation}
In a relevant region of
\begin{equation}
0 < \theta < 2 \tilde{\omega}_R^3 \,, \hspace{0.5cm}
0 < h < 5 \tilde{\omega}_R^2
\,,
\end{equation}
which includes the model region (\ref{model parameter region}),
we numerically checked the time evolution of the homogeneous solution
of eq.(\ref{homogeneous eq in zero friction}).
We observed the power-law increase of the fluctuation.
Indefinite increase of the fluctuation $z(\omega,t)$ as time increase 
implies that the assumed parameter lies either within the unstable band
or on the boundary between the bands.
The linear rise $z(\omega,t) \propto t$, as indicated by our
numerical analysis suggests that the parameter is on the boundary.
The result of the band structure is given in Fig.\ref{h_th}, 
and the relevant parameter 
eq.(\ref{model parameter region}) (corresponding to the crossed point)
is just on the boundary between stable and
1st unstable bands. This is consistent with that
the linear rise, as is expected in the case of the exact Mathiew
function on the boundary \cite{mathiew}.

For this numerical calculation we had to choose
an initial condition for $q_{\rm cl}^{(0)}$. We took several
conditions for investigating band structures, and we always obtained the
same result of the linear power-law increase.

\begin{figure}[t]
  \begin{center}
    \includegraphics[height = 5cm,clip]{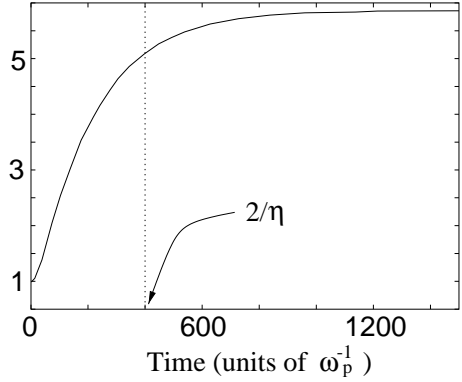}
  \end{center}
  \caption{\small
  Time evolution of the function $y(t)$, a solution of eq.(\ref{y-eq}).
  The amplitude of the function $y(t)$ with the friction
  $\eta = 0.005\omega_p$ is depicted.
  }
  \label{y}
\end{figure}

We next consider effects of the friction $\eta$. The inclusion of the friction
gives a modification of eq.(\ref{homogeneous eq in zero friction}) to
\begin{eqnarray}
  \ddot{z}_h(t)
  +
  [h + 2\theta\, ( q_{{\rm cl}}^{(0)} )^2] z_h(t)
  +
  \eta \dot{z}_h(t)
  = 0~,
\end{eqnarray}
The effect of the friction term $\eta \dot{z}_h(t)$ is studied
by changing the variable $z_h(t)$ to $y(t)$, 
\begin{eqnarray}
  y(t) = z_h(t) e^{\eta t/2}~.
\end{eqnarray}
Then the equation for $y(t)$ becomes
\begin{eqnarray}
  \ddot{y} + [(h - \eta^2/4) + 2\theta\, ( q_{{\rm cl}}^{(0)} )^2] y
  =
  0~.
\label{y-eq}
\end{eqnarray}
On the other hand, the equation for $q_{{\rm cl}}^{(0)}$ is
\begin{eqnarray}
  \ddot{q}_{\rm cl}^{(0)}
  +
  V'(q_{\rm cl}^{(0)})
  +
  \eta \dot{q}_{\rm cl}{(0)}
  =
  0~.
\end{eqnarray}
When the effect of friction for the behavior of the zero-th order
$q_{{\rm cl}}^{(0)}$ is small,
the equation for $y$ is nearly of the parametric resonance type.
In this case then, the behavior of $z(\omega, t)$ is rather simple;
it is a product of the linearly rising function ($y(t)\sim t$)
and the exponentially decreasing function ($e^{-\eta t/2}$). 
The rate of the exponential decrease is $\eta/2$, hence
this product function has a maximum at a time around $2/\eta$.
This seems essentially what we observe in numerical computation for
$z(\omega, t)$.

With the presence of the friction the function $y$ approaches the behavior
of simple harmonic oscillator as time increases since 
$q_{{\rm cl}}^{(0)} \rightarrow 0$ by friction. 
This means, as seen in Fig.\ref{y}, 
that the linear increase of $y$ is saturated at
the time of order $2/\eta$. The net effect of the friction is simply the
minor change of a coefficient of $A/\eta$, where $A$ is close to 1.7.

\begin{figure}[t]
  \begin{center}
    \includegraphics[height = 5.5cm,clip]{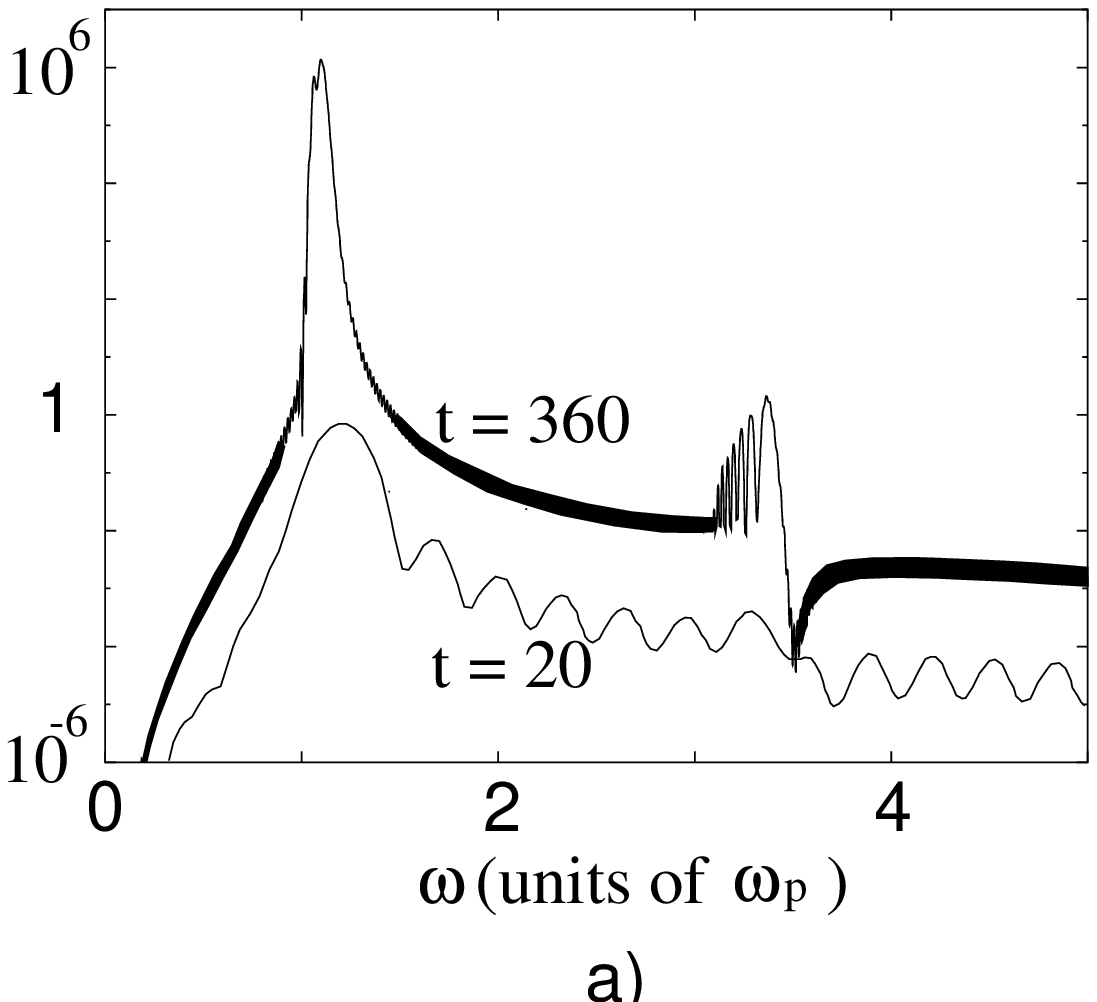}
    \qquad\qquad
    \includegraphics[height = 5.5cm,clip]{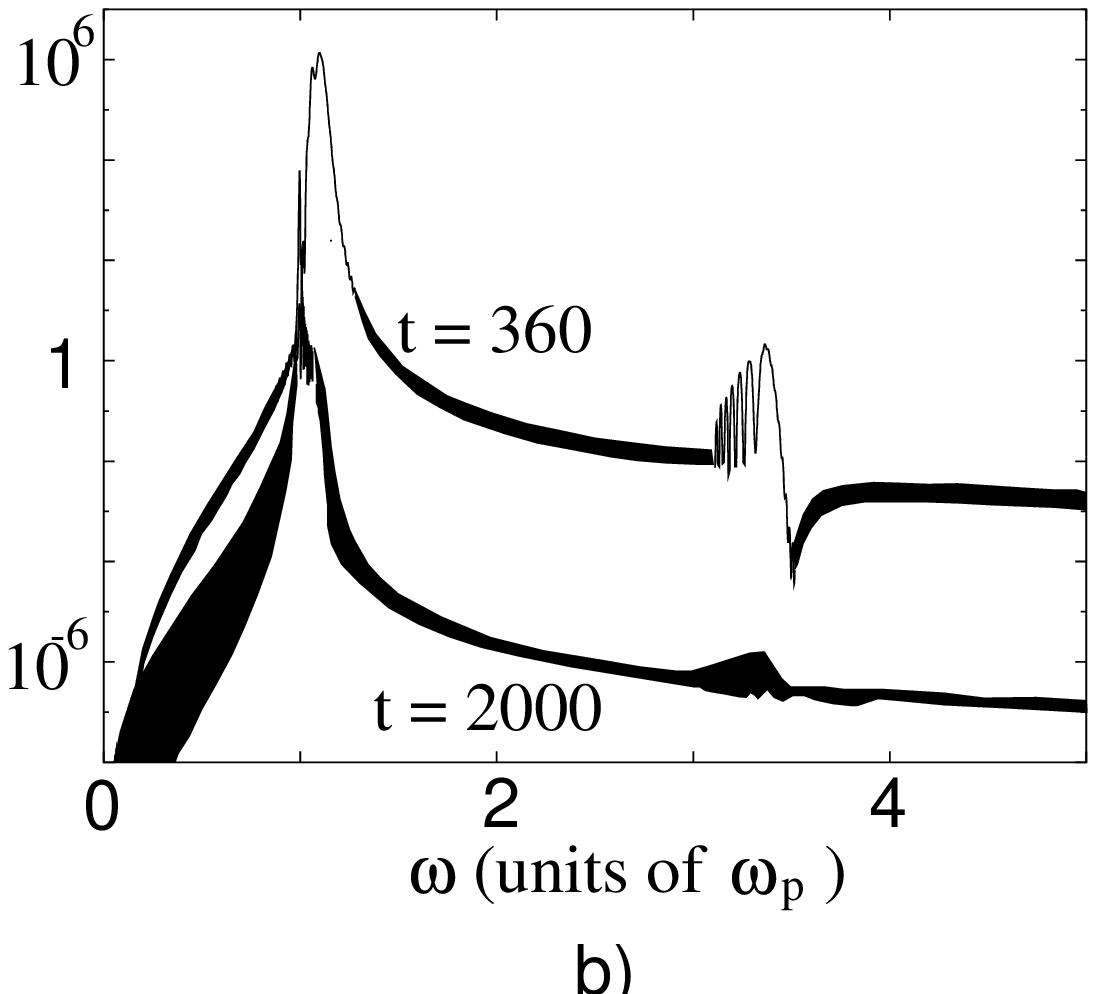}
  \end{center}
  \caption{\small
  Examples of $\omega$ dependence of $z(\omega,t)$.
  The quantity $\omega_p |z(\omega, t)|^2 +
  |\dot{z}(\omega, t)|^2/\omega_p$ both a) at early and b) late 
  times are shown. The unit of time is $\omega_p^{-1}$.
  }
  \label{w-dep}
\end{figure}

\begin{figure}[t]
  \begin{center}
    \includegraphics[height = 5.5cm,clip]{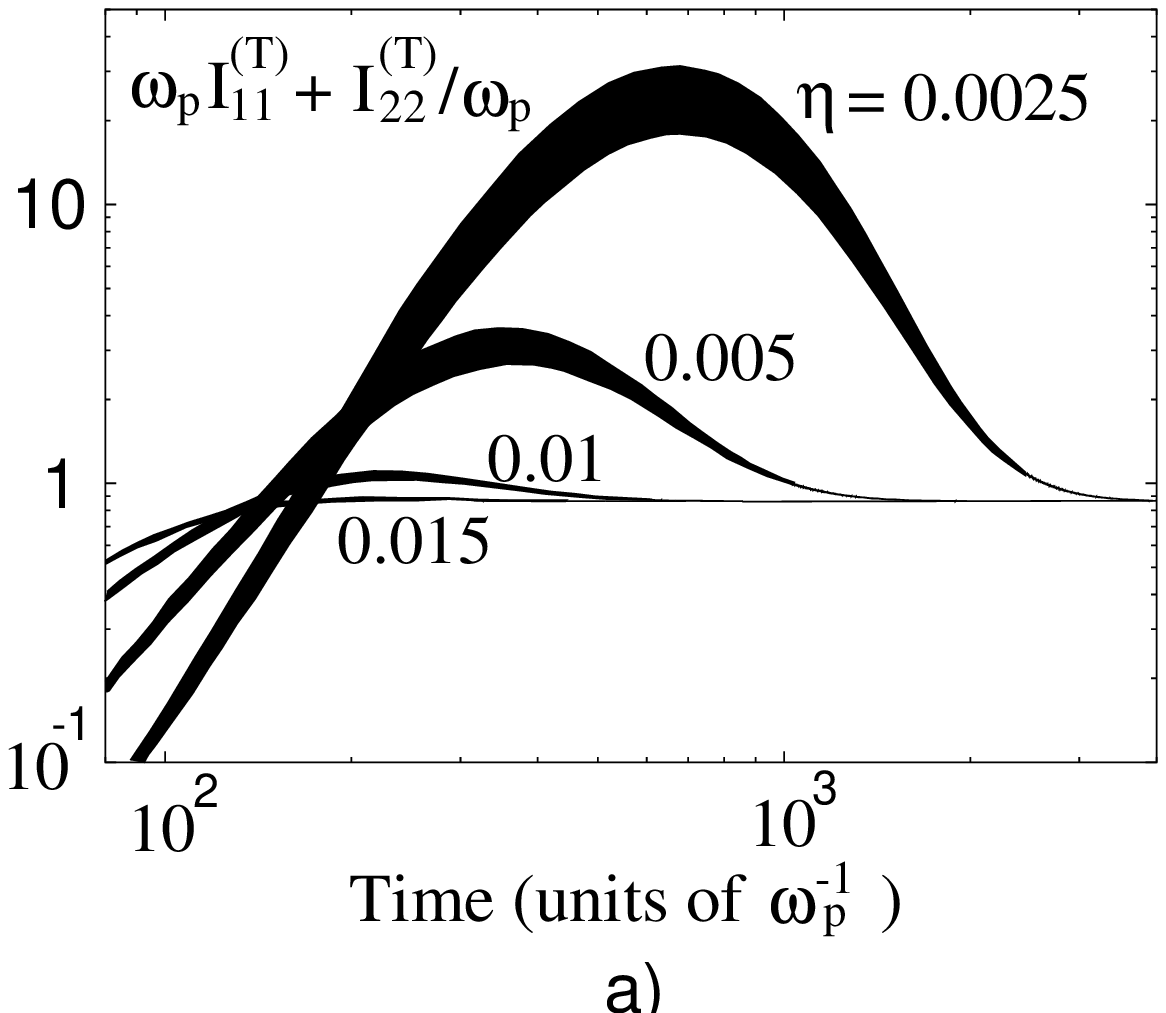}
    \qquad~~~
    \includegraphics[height = 5.5cm,clip]{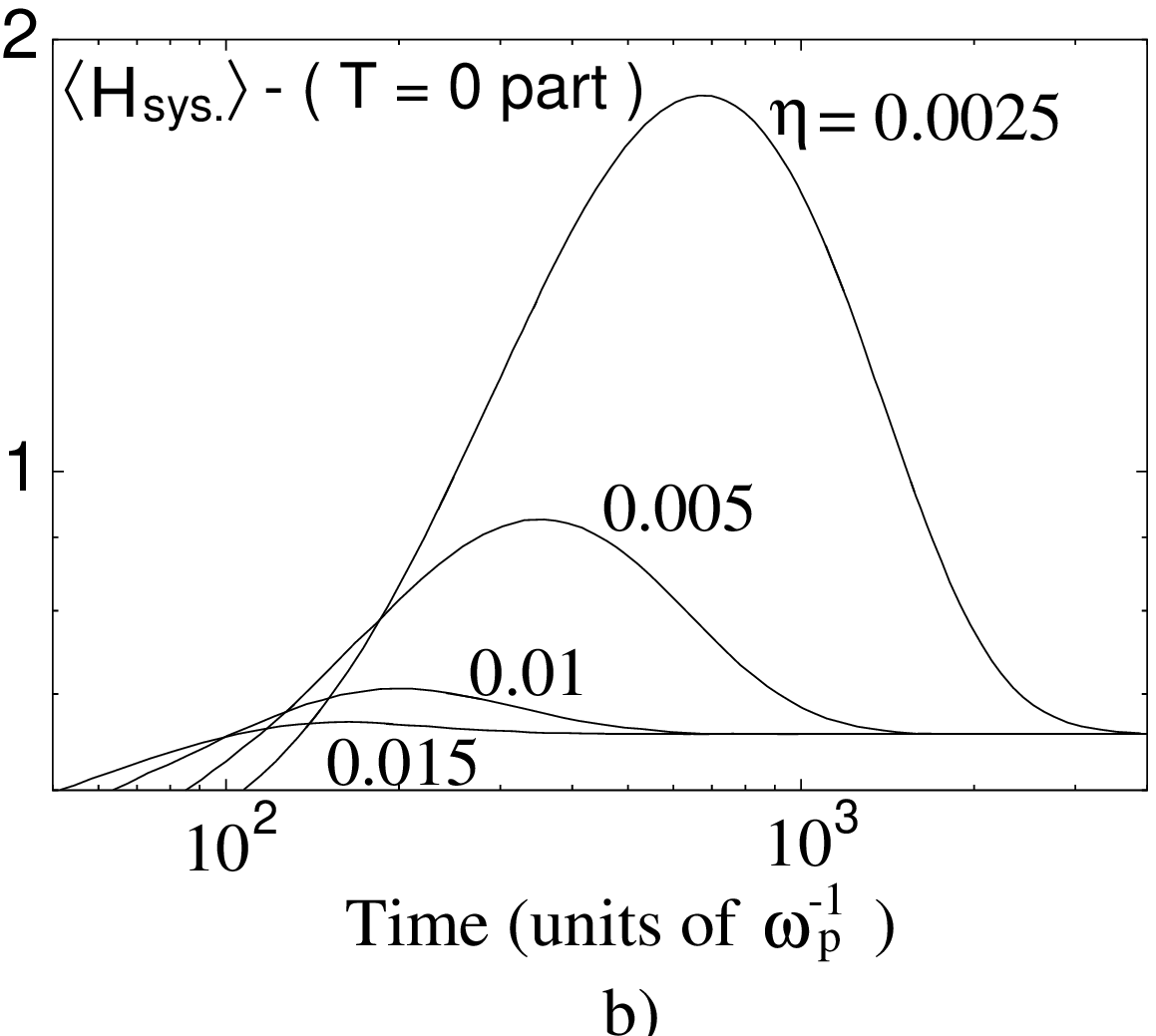}
  \end{center}
  \caption{\small
  Time evolution of a) the fluctuation 
  $\omega_p I^{(T)}_{11} + I^{(T)}_{22}/\omega_p$
  and
  b) the subsystem energy $\langle H_{sys.} \rangle$
  - (T = 0 part). 
  A temperature $T = 0.5\omega_p$ 
  is taken with four cases of the friction $\eta$
  (in the unit of $\omega_p$).
  }
  \label{IandHsys}
\end{figure}

The effect of the presence of the inhomogeneous term in 
eq.(\ref{inhomogeneous fluc eq}) is as follows.
This term acts as an external force, and gives a resonance effect when the
frequency $\omega$ is close to the inherent frequency of the system.
We illustrate this point.
In Fig.\ref{w-dep} $\omega$-dependence  of $\omega_p |z(\omega, t)|^2 + 
|\dot{z}(\omega, t)|^2/\omega_p$ at a particular time
is shown, and one observes, besides the simple resonance, 
higher modes due to the nonlinear effect.
Without the strong resonance effect like this one, 
the fluctuation $z(\omega \,, t)$ is very small, 
since we have the initial condition,
$z(\omega \,, 0) = 0 = \dot{z}(\omega \,, 0)$.
Thus, the environment acts here as a force of resonant kick from
the zero amplitude, which then
makes possible the parametric resonance effect to work, 
typically very
important for large amplitude oscillation.
In Fig.\ref{w-dep}a), the growth of 1st ($\omega \sim \omega_p$) 
and 2nd ($\omega \sim 3\omega_p$) mode is shown.
Other higher modes (3rd, 4th, $\cdots$) are also amplified in 
the same manner as in the 2nd mode case at $t \sim 360\omega_p^{-1}$. 
On the other hand, these higher modes disappear
at late times, and only the first mode remains as seen in Fig.\ref{w-dep}b).
Thus, the growth of higher modes is a finite time effect.

We now discuss the behavior of the frequency integrated 
fluctuation $I(t)$ in eqs.(\ref{fluctuation 11})-(\ref{fluctuation 12}).
The increase of the fluctuation due to the parametric resonance gives 
an interesting time evolution of $I$.
In Fig.\ref{IandHsys}a), the time evolution of the quantity 
$\omega_p I^{(T)}_{11} + I^{(T)}_{22}/\omega_p$ is shown for
a few values of the friction $\eta $.
This function depends on the initial value, $q_i \,, p_i$, for which
we took a typical value in the ground state at the potential minimum.

What is the physical picture behind the resonance enhancement ?
We would like to present a suggestive interpretation.
We first derive a relation between the fluctuation here and the 
subsystem energy defined by
\begin{eqnarray}
\langle H_{sys.} \rangle
  &\equiv&
  \langle \frac{p_{\rm cl}^2}{2}
  +
  \frac{\omega_p^2}{2} q_{\rm cl}^2 \rangle ~.
\end{eqnarray}
By making the approximation on $q_{\rm cl}$ and $p_{\rm cl}$,
namely taking up to linear terms of the initial environment value
as used in eq.(\ref{expansion of cl sol}),
this equation is reduced to 
\begin{eqnarray}
  \langle H_{sys.} \rangle
  &=&
  ({\rm T~=~0~~part})
  \nonumber \\
  &+&
  \frac{1}{2}
  \int^\infty_0 \frac{dp_i dq_i}{2\pi} f^{(i)}_W(q_i,p_i)
  \left\{
    I_{22}^{(T)}(q_i,p_i;t)
    +
    \omega_p^2 I_{11}^{(T)}(q_i,p_i;t)
  \right\}~.
  \label{def h-sys}
\end{eqnarray}
Thus, the temperature dependent part of the subsystem energy is
given by the fluctuation averaged over initial values.
In short, the subsystem energy is driven by the fluctuation.
If initial values for $\langle q_{\rm cl}^2(0) \rangle$, 
and $\langle p_{\rm cl}^2(0) \rangle$ are small (in the $\omega_p = 1$ unit),
the zero-temperature part ($T = 0$) part is close to $0.5 \omega_p$ 
(zero-point fluctuation of the system).
In Fig.\ref{IandHsys}b) the quantity $H_{sys} - $ (T = 0 part)
is shown, 
taking as the initial state the ground state of the harmonic
oscillator at the potential bottom.
The ground state we use is characterized by
\begin{eqnarray}
f^{(i)}_W(q_i,p_i) = 
\frac{1}{\pi}
\exp
\left[
   -\omega_p q_{i}^2 - \frac{p_{i}^2}{\omega_p}
\right]
\,.
\label{initial subsystem wigner fn}
\end{eqnarray}
The time at which the maximum of $H_{sys}$ occurs is $1.7/\eta$, 
which is the same as for that of the fluctuation, thus
indicating a close correlation between the fluctuation and
our subsystem energy $\langle H_{sys.} \rangle$.
We shall have more to say on an important role
of $\langle H_{sys.} \rangle$ towards the end of the next section.

Implication to the tunneling rate seems obvious; 
the parametric resonance we observe here 
enhances excitation to higher energy states
at the potential region left to the barrier top
in the quantum mechanical terminology.
This would then enhance the tunneling rate from the metastable state
of the potential well.
We shall demonstrate this in the following section.

%%%%%%%%%%%%%%%%%%%%%%%%%%%%%%%%%%%%%%%%%%%%%%%%%%%%%%%%%%%%%%%%%%%%%%

\vspace{1cm}
\lromn 5 \hspace{0.2cm} {\bf Tunneling rate formula}
\vspace{0.5cm}

Energy eigenstates are special among pure quantum states,
since they time-evolve only with phase factors.
Thus, if one takes for the projection $|n \rangle \langle n |$ in 
eq.(\ref{def density matrix})
the energy eigenstate $|n \rangle$ of the total Hamiltonian,
the density matrix does not change with time.
On the other hand, if one takes the energy eigenstate
of the subsystem Hamiltonian, then the density matrix
changes solely due to the environment interaction.
It is thus best to use a pure eigenstate of, or its superposition of, 
the subsystem Hamiltonian, when one wishes to determine how the tunneling
rate is modified in thermal medium.

We thus perform projection onto the energy eigenstate of the
tunneling system.
The projected probability onto energy eigenstate of 
the subsystem Hamiltonian is in general given by
\begin{eqnarray}
&&
\hspace*{-0.5cm}
{\cal W}_{E}(t) = \langle E|\rho ^{(R)} | E \rangle =
\int\,dq\,dq'\,dp\,\varphi _{E}(q)\varphi _{E}^{*}(q')\,
e^{ip(q' - q)}\,f_{W}^{(R)}(\frac{q + q'}{2} \,, p \; ; t)
 \,,
\label{projected probability} 
\end{eqnarray}
where $\varphi _{E}(q)$ is the wave function of eigen energy $E$.

Our basic idea is to limit application of this formula to the left region
of the metastable state and to use the semiclassical formula
for the Wigner function \\
$f_{W}(q \,, p \; ; t)$ derived in the preceding section. 
Explicit $p$ integration in $(\ref{projected probability})$ gives
\begin{eqnarray}
&&
{\cal W}_{E}(t) = 
\int\,dq\,dq'\,\frac{dq_{i}dp_{i}}{2\pi }
\,\varphi _{E}(q)\varphi _{E}^{*}(q')\,
f_{W}^{(i)}(q_{i} \,, p_{i})\,\frac{1}{\sqrt{2\pi I_{11}}}\,
\nonumber \\ &&
\hspace*{0.5cm} 
\exp \left[ \,- \,\frac{1}{2I_{11}}\,\left( \frac{q + q'}{2} -
 q_{{\rm cl}}^{(0)}(t) \right)^{2} - \frac{{\rm det}\,I}{2 I_{11}}\,
 (q - q')^{2} 
 \right.
\nonumber \\ &&
\hspace*{0.5cm} 
\left.
 +\, i(q' - q)\,
 \left( \,p_{{\rm cl}}^{(0)}(t) + \frac{I_{12}}{I_{11}}\,
 (\frac{q + q'}{2} - q_{{\rm cl}}^{(0)}(t))\,\right)\,\right]
 \,.
\label{def of projection} 
\end{eqnarray}
The tunneling probability in medium is then obtained by multiplying
the well-known formula of the barrier penetration factor.

Unless the energy eigenstate near the barrier top is important
for computation of the tunneling rate, one may approximate
the wave function in the left region to $q = q_B$ by that of the
harmonic oscillator at the potential bottom, $q = 0$.
The wave function $\varphi _{E}(q)$ is thus approximated by 
using the Hermite polynomial $H_n$ as
\begin{eqnarray}
&&
\varphi _{E}(q) = (\frac{\sqrt{\omega _{p}}}{\sqrt{\pi }2^{n}n!})^{1/2}\,
H_{n}(\sqrt{\omega _{p}}q)\,e^{-\,\omega _{p}q^{2}/2}
 \,,
\label{wave f of ho} 
\end{eqnarray}
where $\omega_p$ is the shifted frequency at the local potential
minimum, eq.(\ref{pole position}).

To compute the projected probability ${\cal W}_{n}(t)$,
we introduce the generating function;
\begin{eqnarray}
&&
{\cal W}(t \,, z) = \sum_{n = 0}^{\infty}\,{\cal W}_n(t)z^n
\,.
\label{generating f for projected prob} 
\end{eqnarray}
The following formula of the Hermite polynomial,
the Mehler's formula \cite{bateman},
\begin{eqnarray}
&&
\sum_{n= 0 }^{\infty }\,\frac{(z/2)^{n}}{n!}\,H_{n}(x)H_{n}(y)
= (1 - z^{2})^{-1/2}\,\exp \left[ \frac{2xyz - (x^{2} + y^{2})z^{2}}
{1 - z^{2}} \right]
 \,.
 \label{Mehler's formula}
\end{eqnarray}
is useful.
From eqs.(\ref{def of projection}), (\ref{wave f of ho}), 
(\ref{Mehler's formula}), one has Gaussian integrals for $q$ and $q'$,
resulting in
\begin{eqnarray}
&&
\hspace*{-0.5cm}
{\cal W}(t \,, z) = 
\int\,\frac{dq_{i}dp_{i}}{2\pi }\,f_{W}^{(i)}(q_{i} \,, p_{i})\,
\sqrt{\frac{1}{{\rm det}\; \tilde{A}}}\,
\exp \left[ \,-\, \frac{1 - z}{2\,{\rm det}\; \tilde{A}}\,
(q_{{\rm cl}}^{(0)} \,, \,p_{{\rm cl}}^{(0)})\,\tilde{A}\,
\left( 
\begin{array}{c}
q_{{\rm cl}}^{(0)}  \\
p_{{\rm cl}}^{(0)}
\end{array}
\right)\,\right]
\,,
\nonumber \\ &&
\label{generating f for projection} 
\\ &&
\hspace*{0.5cm} 
\tilde{A} = 
\left( \begin{array}{cc}
(1 - z)(I_{22} - \frac{\omega_p}{2}) + \omega_p & - (1 - z)I_{12}  \\
-(1 - z)I_{12} & (1 - z)(I_{11} - \frac{1}{2\omega_p})
+ \frac{1}{\omega_p }
\end{array}
\right)
\,.
\end{eqnarray}
Quantities $I_{ij}$ are to be computed at finite time $t$.
The important constraint of the probability conservation
\( \:
\sum_{n = 0}^{\infty}\,{\cal W}_n (t) = 1
\: \)
is automatically satisfied due to ${\cal W}(t \,, 1) = 1$.
For precise computation, one expands this generating function
to powers of $z$, and identifies the projected probability
${\cal W}_{n}(t)$ by (\ref{generating f for projected prob}).

We next clarify the relation,
\begin{eqnarray}
&&
\sum_{n = 0}^{\infty}\,E_n \,{\cal W}_n(t) = \langle H_{sys.} \rangle
\end{eqnarray}
where $\langle H_{sys.} \rangle$ is calculated by (\ref{def h-sys}).
This relation was numerically checked to be correct, as might
be expected.
We have no analytic proof of this relation, although there is
no doubt on the correctness of this relation.

All numerical calculations were performed by using the concrete example
of the tunneling potential eq.(\ref{potential}) throughout this
section. We assume the subsystem is in the ground state at the initial
moment, and use eq.(\ref{initial subsystem wigner fn}) as the initial
Wigner function. Some results were compared with that of the Harmonic
approximation discussed in the last of the section III. 
The effect of the non-linear term is clarified by this comparison,
because the term is neglected in the Harmonic approximation.

\begin{figure}[t]
  \begin{center}
    \includegraphics[height = 5.5cm,clip]{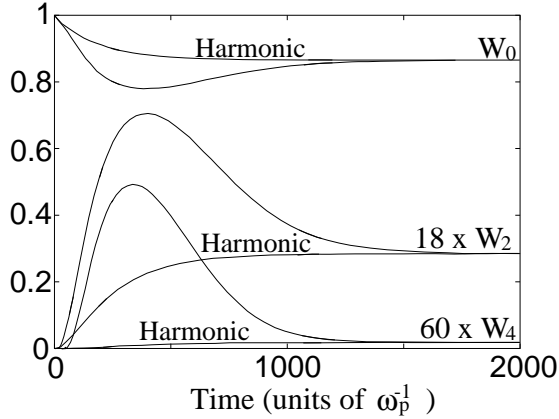}
  \end{center}
  \caption{\small
  Time evolution of the projected probability for the ground, 
  the second excited, and the forth excited state.
  The case of $T = 0.5\omega_p$ and $\eta = 0.005\omega_p$ is depicted.
  For comparison the result in the harmonic approximation is shown.
  }
  \label{Wn}
\end{figure}

In Fig.\ref{Wn}, a few examples of the projected probability
are shown, assuming the initial ground state; ${\cal W}_n (0) = \delta_{n 0}$. 
It is clearly observed that excitation to higher energy levels
is correlated to evacuation of the ground state.
The infinite time limit of ${\cal W}_n (\infty)$ coincides with
the result of the harmonic approximation at the potential bottom,
\begin{eqnarray}
{\cal W}_n (\infty) = 
(1 - e^{-\beta\omega})e^{-n\beta\omega} + {\cal O}(\eta)
\,,
\end{eqnarray}
as is expected.
We note that the subsystem thermalizes at the infinite time
despite that it was initially in the ground state of a harmonic 
oscillator, hence a non-equilibrium state.

The tunneling rate $\Gamma (t)$ per unit time is roughly
the tunneling probability
\( \:
\times \frac{\omega _{p}}{2\pi } \,, 
\: \)
since $\omega _{p}/2\pi  =$ 1/period for
the periodic motion in the left potential well.
Thus, 
\begin{eqnarray}
&&
\Gamma (t) = \frac{\omega _{p}}{2\pi }\,\sum_{n= 0}^{\infty }\,
{\cal W}_{n}(t) \,|T(E_{n})|^{2}
\,, 
\label{tunneling rate}
\\ &&
|T(E)|^{2} \approx 
\exp \left[\,
-\,2\,\int_{q_{1}(E)}^{q_{2}(E)}\,dx\sqrt{2( \tilde{V}(x) - E)}
\,\right]\,\theta (V_{h} - E)  + \theta (E - V_{h}) \,.
\end{eqnarray}
In computing the tunneling rate we use a modified potential
$\tilde{V}$ which is obtained from $V$ by replacing the
curvature parameter $\omega _{R}$ by $\omega _{B}
= \sqrt{\omega _{R}^{2} + \eta ^{2}/4} - \eta /2$.
The reason for this is that one should compare the tunneling rate
with and without the environment effect by using effectively the same
values of parameters for the two cases.
This modification explains the bulk of the environment effect,
the environment suppression factor proposed by \cite{caldeira-leggett 83},
as shown in \cite{my 00-1}.
In equations above 
$q_{i}(E)$ are turning points separating the subbarrier region.

\begin{figure}[t]
  \begin{center}
    \includegraphics[height = 5.7cm,clip]{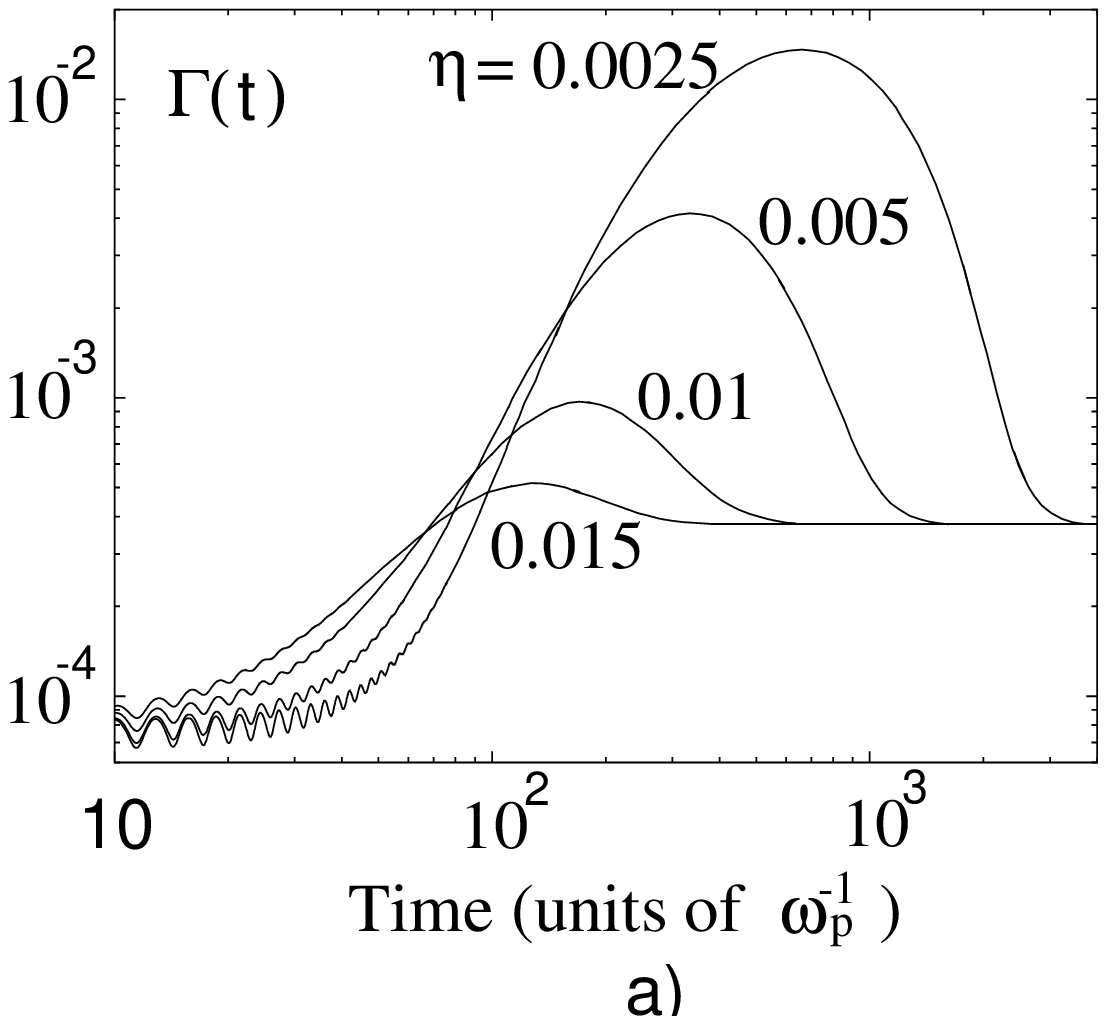}
    \qquad~~~
    \includegraphics[height = 5.cm,clip]{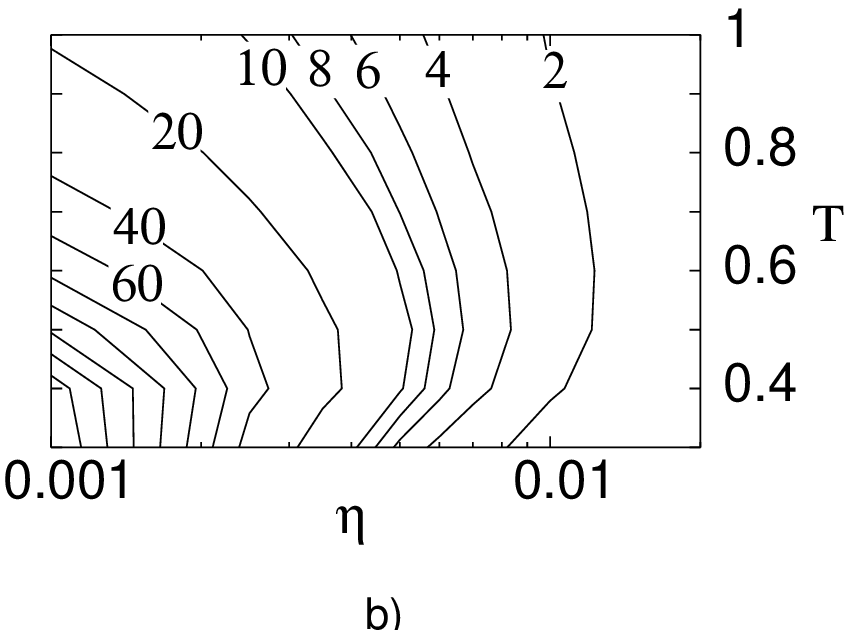}
  \end{center}
  \caption{\small
  a) Time evolution of th tunneling rate $\Gamma(t)$. 
  A temperature $T = 0.5\omega_p$ is
  taken with four cases of the friction $\eta$.
  b) Ratio of the maximal tunneling rate at $t \sim {\cal O}(2/\eta)$
  to the $\Gamma(\infty)$ (a traditional one) is depicted 
  as a contour map in the $\eta-T$ plane.
  }
  \label{GandGMax}
\end{figure}

In Fig.\ref{GandGMax}a) the tunneling rate $\Gamma (t)$ is shown for a few
values of the friction.
The tunneling rate is observed to become maximal at the time
$t \sim 1.7/\eta$.
A salient feature, and the most important result of the present work,
is that the tunneling rate is enhanced around the time of order $2/\eta $,
which is caused by non-linear resonance effects.
The enhanced tunneling rate at $t \approx O(2/\eta )$ is clearly
related to the power-law rise of the fluctuations $I_{ii}$,
as observed in our previous paper \cite{my 00-3} and as already explained.
The maximal enhancement factor for the tunneling rate is plotted in
Fig.\ref{GandGMax}b), relative to the harmonic approximation.
A large enhancement is obtained for small $\eta$ and small $T$ values.
In the harmonic approximation 
an almost monotonic behavior of time evolution is evidently
observed, thus the rate being given by $\Gamma(\infty)$.
We may summarize what happens, in the following way.
The non-linear resonance inherent to the tunneling potential
excites higher modes that have larger tunneling probabilities.
This effect is however terminated by the friction caused
by environment interaction, which is the origin of
the final decrease of the tunneling rate at large times.

We finally study at a quantitative level
correlation between the tunneling rate $\Gamma(t)$ and 
the tunneling system energy $H_{sys}(t)$.
In order to evaluate the correlation quantitatively, 
we define the correlation coefficient $r$ used in statistics;
\begin{eqnarray}
  r = \frac{\langle \Gamma H_{sys}\rangle
            - \langle \Gamma\rangle \langle H_{sys}\rangle}
           {\langle\Gamma\rangle\langle H_{sys}\rangle},
  \qquad
  \langle A \rangle
  \equiv
  \frac{1}{\rm Time}\int_0^{\rm Time} dt~A(t)
  \,.
\end{eqnarray}
The correlation coefficient $r$ was numerically calculated and
we obtained results that can be summarized by $r = 0.99 \sim 1$ in the 
parameter region of temperature and friction, $0.2 \leq T/\omega_p
\leq 1.0$ and $0.001 \leq \eta/\omega_p \leq 0.02$. 
This proves the correlation between the two quantities.

\begin{figure}[t]
  \begin{center}
    \includegraphics[height = 5.7cm,clip]{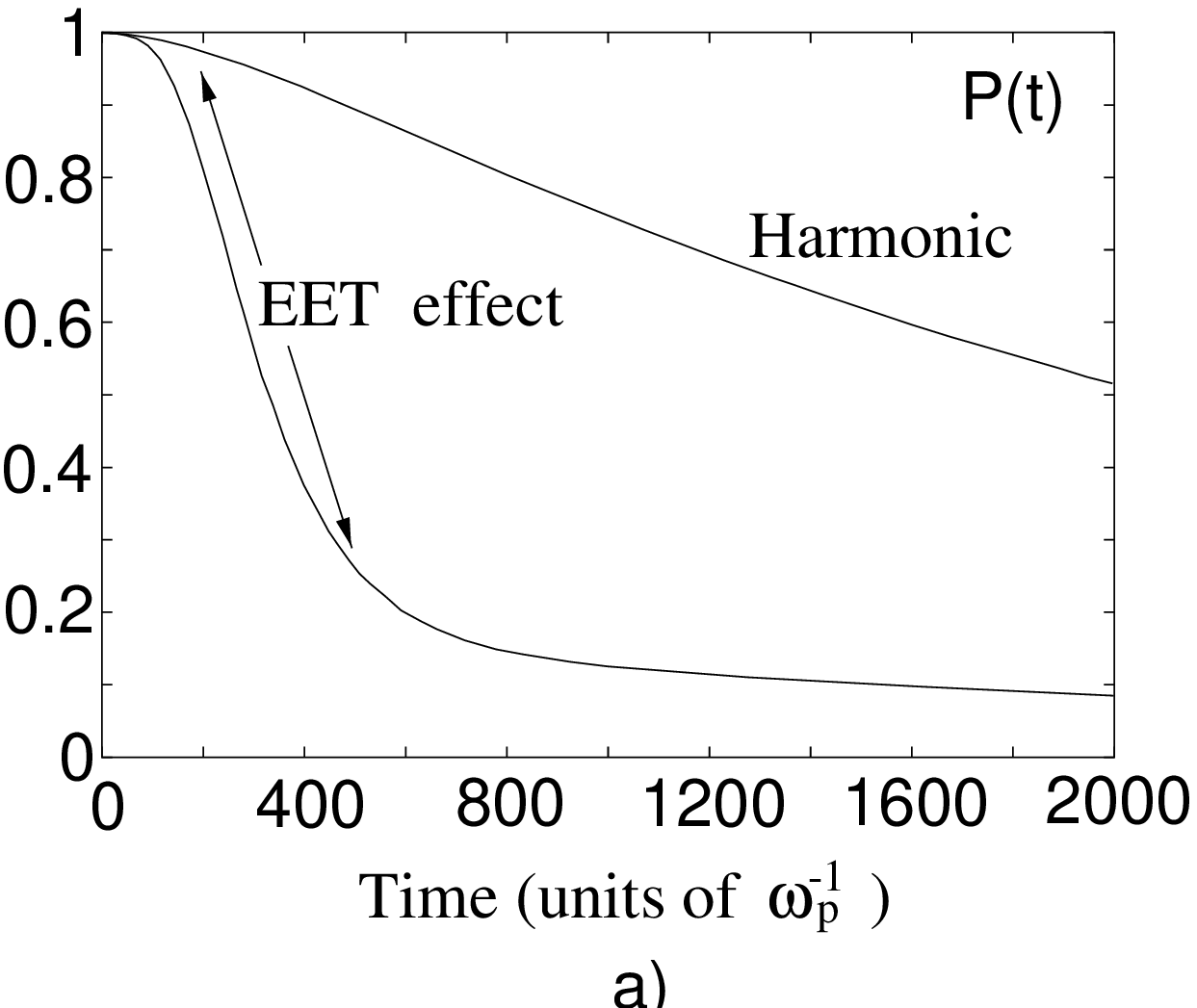}
    \qquad~~~
    \includegraphics[height = 5.cm,clip]{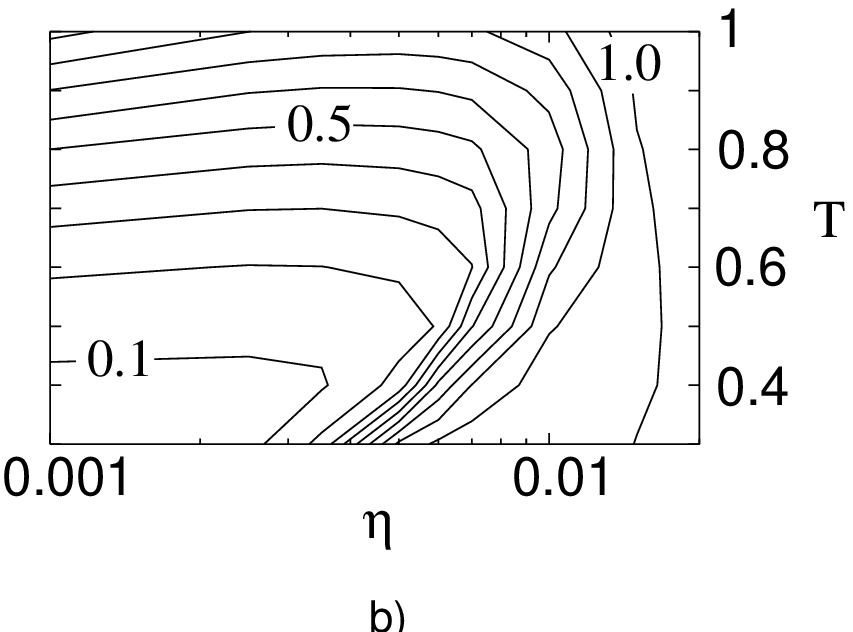}
  \end{center}
  \caption{
  a) Time evolution of the survival probability $P(t)$.
     The case of $T = 0.5\omega_p$ and $\eta = 0.005\omega_p$ 
     is depicted. For comparison the result in the harmonic approximation
     is shown.
  b) Ratio of the lifetime as defined in eq.(\ref{effective life}) 
     to the traditional one is
     depicted as a contour map in the $\eta - T$ plane.
  }
  \label{PandLT}
\end{figure}

We next turn to the discussion of the survival probability,
the probability that the metastable state remains in the same initial
state.
 This quantity  $P(t)$ is defined by
\begin{eqnarray}
  \dot{P}(t) = -\Gamma(t) P(t), \qquad P(0) = 1 \,.
\end{eqnarray}
This gives
\begin{eqnarray}
  P(t) = \exp
  \left[
    -\int^t_0 dt'~\Gamma(t')
  \right] \,.
\end{eqnarray}
In Fig.\ref{PandLT}a) 
the survival probability is depicted along with the result of
the harmonic approximation of the potential bottom. 
The effect of EET is unquestionable.

A traditional definition of the lifetime $t_{\rm trad}$ 
which many authors use is
\begin{eqnarray}
  t_{\rm trad} = \Gamma^{-1}(\infty) \,,
\end{eqnarray}
while in the case of time dependent tunneling
one may effectively use $t_{\rm LT}$ given by the e-holding time,
\begin{eqnarray}
  P(t_{\rm LT}) = e^{-1} \,.
\label{effective life}
\end{eqnarray}
In Fig.\ref{PandLT}b) a contour plot of $t_{\rm LT}/t_{\rm trad.}$ is shown
in the parameter ($\eta \,, T$) space.
For small values of $\eta$ and $T$ a substantial reduction of the lifetime
$t_{\rm LT}$ is observed, which means the early termination of the tunneling.

We hope that we have convinced the reader of the power and the usefulness
of EET, although our idea is vindicated by detailed numerical analysis
which is necessarily limited to special cases.

%%%%%%%%%%%%%%%%%%%%%%%%%%%%%%%%%%%%%%%%%%%%%%%%%%%%%%%%%%%%%%%%%%%%%%%%%%

\vspace{0.5cm}
\lromn 6 \hspace{0.2cm} {\bf Discussion}
\vspace{0.5cm} 

There may be many applications of environment-driven excited tunneling (EET) 
presented in this paper.
We shall only mention two possible applications to cosmology, 
an enhancement mechanism of electroweak baryogenesis and a possible
resurrection of the old inflationary scenario of the GUT phase transition .
Furthermore in the present discussion we shall focus on
one aspect of the baryogenesis condition; the out-of-equilibrium
condition \cite{sakharov}, \cite{my}.
Needless to say, there are many problems to be solved in actual
application of these ideas to realistic models.

Electroweak baryogenesis is expected to occur if the 
electroweak phase transition is of a strong first order. 
The first order phase transition
usually proceeds via the bubble formation. Inside the bubble the true
electroweak broken phase is realized in which the baryon number is 
effectively conserved.
On the other hand, outside the bubble the universe remains in the high 
temperature symmetric 
phase in which the baryon number is not conserved via sphaleron related
processes\cite{krs}. This mismatch inside and outside the bubble 
gives rise to a possibility of electroweak baryogenesis, 
although details of a promising mechanism are yet to be worked out.
Thus, the strong first order of the phase transition has been considered to be
a requisite for the out-of equilibrium condition necessary 
for the baryogenesis.

In the ordinary circumstance, namely in the standard model, 
the strong first order phase transition
requires a small Higgs mass, which is inconsistent with observation.
Along with the small CP violation effect of Kobayashi-Maskawa theory,
this excludes \cite{ew-bgenesis} the possibility \cite{fs} 
of the standard model as a viable model for baryogenesis.
Thus, a more complicated model such as two Higgs doublet model needs
to be invoked for the electroweak baryogenesis.
Regardless of whether an extended Higgs model is needed, we would like to
point out that the 
environment-driven excited tunneling (EET) found in this paper
has a chance of enhancing the out-of equilibrium condition necessary for
the baryogenesis.

Although not much discussed in the literature, 
there is an important factor to consider for
the out-of equilibrium condition. 
This is the factor $\Gamma t_H$, where $\Gamma$ 
is the nucleation rate of the bubble and $t_H = 1/H$ is the Hubble time.
In the usual scenario of the first order phase transition 
the effectiveness of electroweak baryogenesis is in proportion to this
quantity. 
The nucleation rate $\Gamma$ in the usual estimate without EET is exponentially
suppressed by the potential barrier, and the factor $\Gamma t_H$ is much less
than 1, which means that the first order phase transition is never completed by
a merger of nucleated bubbles. 
Under this circumstance it is expected that the phase transition is
terminated by proceeding to disappearance of
the local minimum of the symmetric phase so that the condition for the first
order phase transition is not met. 
The effectiveness of the out-of equilibrium condition is then reduced by
a factor proportional to the quantity $\Gamma t_H$.

On the other hand, with the aid of EET, the factor $\Gamma t$ may become of
order unity at some finite time, 
and the phase transition may be completed much prior to the
Hubble time, as discussed in previous sections.
One should check in comparison to the Hubble time the new time scale $2/\eta$ 
which is roughly the time at the maximum for the tunneling rate.
The factor $\Gamma t$ thus may become of order unity at $t \sim 2/\eta$.
In this case there is no reduction factor like $\Gamma t_H$.

We shall check numerical relation between two time scales; $t_H$ and $2/\eta$.
The size of the friction $\eta$ in the standard model 
is of the order of the Higgs mass $m_H$ 
times some power of coupling factors, while the Hubble rate is given by
\begin{eqnarray}
H = 1/t_H = \sqrt{\frac{4\pi^2}{45}N}\frac{T^2}{m_{pl}} \,, 
\qquad N = 106.75
\,.
\end{eqnarray}
Calculation, as summarized in Appendix, indicates that
\begin{eqnarray}
\eta \simeq \frac{y_t^2}{32\pi}\omega_p\tanh(\frac{\omega_p}{4T}), 
\qquad
\omega_p^2 = V''(0) \simeq 0.2\times(T^2 - m_H^2 - 0.04 v^2)
\,,
\end{eqnarray}
where $y_t$ is the top Yukawa coupling, and
$V(\phi)$ is the effective potential of the standard model at finite
temperature, with  $v$ the vacuum expectation value of the Higgs field.
Thus, the Hubble rate is much smaller than $\eta$ by the inverse
Planck factor since $\frac{m_H}{m_{pl}} \ll y_t ^2$.

One may expect that even in the case of weak first order phase
transition the quantum tunneling enhanced by EET may give rise to
a sufficient condition for the out-of equilibrium in the early
phase of the phase transition.
Much remains to be seen with this regard.

Another interesting effect of EET concerns resurrection of the old inflationary
scenario of the GUT phase transition \cite{old inflation}. 
The old inflationary scenario
was once rejected because the bubble formation proceeds too slowly,
hence leaving a very inhomogeneous universe \cite{problem of old inflation}.
The essential reason of this problem, called graceful exit problem, is that
the exponentially suppressed tunneling rate is too small such that
bubbles of true vacuum do not merge sufficiently to give rise to a homogeneous
universe which the inflationary scenario intends to derive.
Our mechanism of enhanced tunneling may help this situation.
The problem must however be addressed quantitatively; one has to obtain
an early termination of the first order phase transition to promote
merger of bubbles, yet has to obtain an sufficient inflation to solve
intended problems.
We leave this quantitative study to future.

\vspace{2cm}

\begin{center}
{\bf Acknowledgments}
\end{center}

This work is supported in part by the Grant-in-Aid for Science Research,
Ministry of Education, Science and Culture, Japan (No.13640255 and
No.13135207)

\newpage
\begin{center}
\begin{large}
{\bf Appendix}
\end{large}
\end{center}

\vspace{0.7cm}
{\bf A-1 Effective potential and critical temperature}
\vspace{0.2cm}

\hspace{0.2cm}
We shall first give  the finite temperature effective potential of 
the standard model at 1-loop order, 
assuming that the Higgs mass is much smaller than other particle masses
of the standard model such as the top quark, the weak boson, thus
$m_H << m_t \,, m_Z \,, m_W$. This is taken as a necessary condition 
for the first order phase transition. This condition however is 
not met in reality. Nevertheless, we take this case as an important
illustration for more complicated cases.

The zero temperature part of the effective potential \cite{coleman-weinberg}
is as follows;
\begin{eqnarray}
  V_0(\phi)
  &=&
  \frac{m_H^2}{8v^2}(\phi^2 - v^2)^2
  +
  2Bv^2\phi^2
  -
  \frac{3}{2}B\phi^4
  +
  B\phi^4\log\left(\frac{\phi^2}{v^2}\right)
  \,,
  \nonumber \\
  B
  &=&
  \frac{3}{64\pi^2v^4}\left(2m_W^4 + m_z^2 - 4m_t^2\right)  
  \simeq -0.005
\,,
\end{eqnarray}
where $v = 246 GeV$ is the vacuum expectation value (VEV) of the Higgs field, 
and $m_a$ is the mass of a particle species $a$.
The finite temperature effect is then given by the free energy
\cite{effective potential in sm},
\begin{eqnarray}
  V_{T}(\phi)
  &=&
  \frac{T^4}{2\pi^2}
  \left[
    6I_B\left(\frac{m_W\phi}{vT}\right)
    +
    3I_B\left(\frac{m_z\phi}{vT}\right)
    -
    6I_F\left(\frac{m_t\phi}{vT}\right)
  \right]
  \,,
  \nonumber \\
  I_{B,F}(a^2)
  &=&
  \int^\infty_0 dr r^2 \log
  \left[
    1 \mp \exp\left(-\sqrt{r^2 + a^2}\right)
  \right]
\,.
\end{eqnarray}
The Higgs contribution has been neglected, due to the assumption of
$m_H \ll $ masses of other particles.

According to \cite{effective potential in sm}, we may make 
the high temperature expansion of the effective potential with
$T \ll v$, to get
\begin{eqnarray}
  V(\phi)
  &=&
  V_0(\phi) + V_{T}(\phi)
  \simeq
  D(T^2 - T_0^2)\phi^2 - ET\phi^3 + \frac{\lambda(T)}{4}\phi^4
  \,,
\end{eqnarray}
where the dimensionless coefficients are given by
\begin{eqnarray}
  D
  &=&
  \frac{2m_W^2 + m_z^2 + 2m_t^2}{8v^2}
  \simeq
  0.2
\,,
  \nonumber \\
  E
  &=&
  \frac{2m_W^3 + m_z^3}{4\pi v^3}
  \simeq
  0.01
  \,,
  \nonumber \\
  T_0^2
  &=&
  \frac{m_H^2 - 8Bv^2}{4D}
  \simeq
  1.25(m_H^2 + (50GeV)^2)
  \,,
  \\
  \lambda(T)
  &=&
  \frac{m_H^2}{2 v^2} 
  -
  \frac{3}{16\pi^2v^4}
  \left(
    2m_W^4 \log\frac{m_W^2}{a_BT^2}
    +
    m_z^4 \log\frac{m_z^2}{a_BT^2}
    -
    4m_t^4 \log\frac{m_t^2}{a_FT^2}
  \right)
\,.
  \nonumber 
\end{eqnarray}
Here $a_B =2 \log 4\pi - 2\gamma \simeq 3.91$, 
$a_F =2 \log \pi - 2\gamma \simeq 1.14$.
The critical temperature is computed from these, to give
\begin{eqnarray}
  T_c^2
  =
  \frac{T_0^2}{1 - E^2/(\lambda(T)D)}
 \simeq \frac{T_0^2}{1 - (8 GeV/m_H)^2}
  \,.
\end{eqnarray}
At the value of $T_c$ two local minima coincide, while
$T_0$ indicates where the symmetric phase of $\varphi = 0$ 
disappears as the local minimum.
It is thus important to note that the condition $T_c > T_0$
for the first order phase transition is met.

\vspace{0.2cm}
{\bf A-2 Friction coefficient in the standard model}

\vspace{0.2cm}
The friction coefficient $\eta$ is computed using 
the method of \cite{dissipation coefficient},\cite{dissipation coefficient 2}.
According to \cite{dissipation coefficient 2}, 
it is possible to identify the two-body bilinear operator of 
standard model fields as the environment variable $Q(\omega)$.
The most dominant contribution to the friction $\eta$
comes from the top loop diagram where the top-Yukawa coupling is of the form,
$y_t \,\phi \bar{t}t /\sqrt{2}$.
Thus, the Langevin equation of motion for the zero-mode Higgs field $\phi_0$ 
becomes
\begin{eqnarray}
  &~&
  \ddot{\phi_0} + V'(\phi_0)
  +
  2\int^t_0 ds \Im \alpha(t - s)\phi_0(s) = F(t)
  \,, 
\label{0-mode eq}
\\
  &~&
  \langle \{F(t),F(s)\}\rangle_{\rm env} = \Re \alpha(t - s)
  \,, \\
  &~&
  \alpha(x_0)
  =
  \frac{y_t^2}{2}
  \int d^3x
  \langle
    \bar{t}t(x)\bar{t}t(0)
  \rangle_{\rm env}
  =
  \int d\omega \frac{r_H(\omega)}{e^{\beta\omega} - 1}e^{i\omega x_0} ~.
\end{eqnarray}
The real-time Green's function 
$\langle\bar{t}t(x)\bar{t}t(0)\rangle_{\rm env}$ 
can be obtained by the analytic continuation of the imaginary-time
Green's function \cite{Fetter-Walecka}, 
and it is easy to calculate the imaginary-time Green's
function by using the Matsubara formalism. 
The necessary spectral weight $r_H(\omega)$ is given by
\begin{eqnarray}
  r_H(\omega) = \frac{y_t^2}{2}\int \frac{d^3p}{(2\pi)^3}\,
  \frac{\omega^2}{4p^2}
  \delta(\omega - 2p)(1 - 2n_p)~ \,,
\label{spectral in sm}
\end{eqnarray}
where $p = |\vec{p}|$ is the momentum magnitude and 
$n_p = (e^{\beta p} + 1)^{-1}$,
assuming that the top quark is massless in the symmetric phase
around $T \approx T_0$.

In the vicinity of the false vacuum, namely $\varphi \approx 0$,
the behavior of the zero-mode Higgs field
exhibits the exponentially dumped oscillation.
The solution of the homogeneous zero-mode Higgs
field equation (\ref{0-mode eq}) is approximately obtained by
\begin{eqnarray}
  \phi_0(t)
  &\simeq&
  2\int^\infty_0 d\omega
  \frac{r_H(\omega)\sin(\omega t)}
       {(\omega^2 - \omega_p^2) + \pi^2 r^2_H(\omega)}
  \nonumber \\
  \nonumber \\
  &\simeq&
  2\int^\infty_0 d\omega
  \frac{r_H(\omega_p)\sin(\omega t)}
       {(\omega^2 - \omega_p^2) + \pi^2 r^2_H(\omega_p)}
  = \sin(\omega_p t)\exp(-\frac{\pi r_H(\omega_p)}{2\omega_p}t)~ \,,
\end{eqnarray}
where $\omega_p$ is the curvature of the false vacuum
($\omega_p^2 = V''(0)$), and the friction 
coefficient in this case is computed from eq.(\ref{spectral in sm}).
The result is
\begin{eqnarray}
  \eta
  &=&
  \frac{\pi}{\omega_p}r(\omega_p)
  =
  \frac{y_t^2}{32\pi}\omega_p\tanh\frac{\beta \omega_p}{4}~ \,.
\end{eqnarray}

%%%%%%%%%%%%%%%%%%%%%%%%%%%%%%%%%%%%%%%%%%%%%%%%%%%%%%%%%
\vspace{1cm} 

\end{document}